\documentclass[preprint,12pt]{elsarticle}

\usepackage{amssymb,amsmath}

\usepackage{multirow}
\usepackage{booktabs}
\usepackage{color}
\usepackage{longtable}
\usepackage{floatrow}
\usepackage[labelsep=space]{caption}
\usepackage{geometry}
\DeclareFloatFont{tiny}{\tiny}
\usepackage{ulem}

\journal{Computers \& Fluids}

\begin{document}

\begin{frontmatter}


\title{Implementation of a discrete Immersed Boundary Method in \textit{OpenFOAM}}

\author[marseille]{E. Constant}
\author[marseille]{C. Li}
\author[marseille]{J. Favier\corref{cor2}}
\ead{julien.favier@univ-amu.fr}
\author[poitiers]{M. Meldi}
\author[marseille]{P. Meliga}
\author[marseille]{E. Serre}
\cortext[cor2]{Corresponding author}

\address[marseille]{Aix-Marseille Universit\'e, CNRS, \'Ecole Centrale Marseille, Laboratoire M2P2 UMR 7340, 13451, Marseille, France}
\address[poitiers]{Institut $P^{\prime}$, CNRS -- Universit\'{e} de Poitiers -- ENSMA,  11 Bd Marie et Pierre Curie, 86962 Futuroscope Chasseneuil Cedex, France}

\begin{abstract}
In this paper, the Immersed Boundary Method (IBM) proposed by \citet{Pinelli20109073} is implemented for finite volume approximations of incompressible Navier-Stokes equations solutions in the open source toolbox \textit{OpenFOAM} version 2.2 (\citet{OpenFOAM}). Solid obstacles  are described using a discrete forcing approach for boundary conditions. Unlike traditional approaches encompassing the presence of a solid body using conformal meshes and imposing no-slip boundary conditions  on the boundary faces of the mesh, the solid body is here represented on the Eulerian Cartesian mesh through an ad-hoc body force evaluated on a set of Lagrangian markers. The markers can  move across the Eulerian mesh, hence allowing for a straightforward analysis of motion or deformation of the body. 
The IBM method is described and implemented in PisoFOAM, whose Pressure-Implicit Split-Operator (PISO) solver is modified accordingly. The presence of the solid body and the divergence-free of the fluid velocity are imposed simultaneously by sub-iterating between IBM and the pressure correction step. This scheme allows for the use of fast optimized Poisson solvers while granting excellent accuracy with respect to the previously mentioned constraints. Various 2D and 3D  well-documented test cases of flows around fixed or moving circular cylinders are simulated and carefully validated against existing data from the literature. The capability of the new solver is discussed in terms of accuracy and numerical performances.
\end{abstract}


\end{frontmatter}


\section{Introduction}
\label{sec::Intro}

The numerical simulation of industrial and environemental flows generally involves complicated aspects such as complex geometries and high Reynolds number flows. An accurate description of these flows can be achieved using efficient numerical modeling. In addition, in many configurations, the fluid may interact with solid structures which requires to relevantly model and implement such interactions within powerful numerical codes. The long-term goal of this work is to develop such a numerical tool using the C++ libraries \textit{OpenFOAM}, released under the GNU Public license (GPL).  Open source CFD codes generally provide efficient coding and optimized tools running on massive parallel computers, plus they provide suitable environments for implementation and rapid dissemination of new algorithms to the users community. 

\textit{OpenFOAM} (\citet{OpenFOAM}) is an extended repository of C++ libraries which allows for the numerical simulation of a wide range of applications.
It has gained a vast popularity during the recent years as the user is provided with existing solvers and tutorials allowing for a quick start to using the code. The software is now extensively used both in academic research (see among others the papers by  \citet{Tabor2010553,Meldi2012, Lysenko13, Komen14}) and for industrial flows analysis (\citet{Selma2014241, Flores13, Gao12}). \textit{OpenFOAM} solvers can also be freely modified to become more efficient, and several papers in the literature deal with the implementation of new numerical techniques or models in \textit{OpenFOAM} (see among others the papers by \citet{Flores13, Towara15, Vuorinen14}). Here, the focus is on addressing efficiently a correct description of flows around obstacles with complex geometries.  In \textit{OpenFOAM}, immersed bodies are primarily accounted by the use of wall-boundary conditions. However, when dealing with complex geometries, this approach leads to significant deformations of the computational mesh. On the one hand, this yields non-negligible numerical errors that are usually difficult to estimate. On the other hand, although body-fitted coordinate systems may yield a well-suited discretization of given geometry (\citet{Ferziger96}), the grid generation may become a prohibitive issue if the geometry varies in time, as is commonly encountered in fluid-structure interaction problems. This clearly stresses the need to develop specific, advanced numerical techniques to address such complex configurations.

An alternative and more recent approach is the Immersed Boundary Method (IBM). A wide spectrum of methods included in this family have proven efficient to simulate complex and moving geometries, such as Lagrangian multipliers (\citet{Glowinski99}), level-set methods (\citet{Cheny10}), fictitious domain approaches and surface (\citet{Peskin1972252}) and volume penalization approaches (\citet{Minguez08, Isoardi10}). The present work, deals with the IBM primarily proposed in the seminal work of \citet{Peskin1977220}, who introduced this method to simulate fluid-structure interactions into a cardio-vascular system (see the late, seminal paper by \citet{peskin2002} for the mathematical foundation). A common feature of all IBM techniques is that the Navier-Stokes equations are discretized over a simple structured Cartesian grid, which significantly improves the computational efficiency and the stability. The complex geometry is then immersed into a larger computational domain, and the boundary conditions are represented by the addition of an ad-hoc body force in the momentum equations. Such a body force is meant to mimick the effect of classical no-slip boundary conditions at the physical surface of the obstacle.

Several improvements and extensions of Peskin's method  have been proposed in the literature. They can be classified into two families, continuous or discrete forcing (\citet{Mittaletal2005}), depending on whether the force is applied on continuous or discretized Navier-Stokes equations. 
The original Peskin's method is an example of continuous forcing method. The fluid is represented on an Eulerian system of coordinate whereas the structure is represented on a Lagrangian one, where markers define immersed solid boundaries. Approximations of the Delta distribution by smoother functions allow to interpolate between the two grids (\citet{Peskin1977220}). Since then, other formulations have been proposed. For instance, the feedback forcing method is based on the idea of driving the boundary velocity to rest ( \citet{beyeretal1992,goldsteinetal1993}). These methods are not sensitive to the numerical discretization, but suffer from limitations due to the use of free constants in their formulation. Moreover, they are also subjected to spurious oscillations and severe CFL restrictions related to stiffness constants (\citet{Mittaletal2005}).  

The direct forcing approach, also termed the discrete approach, aims at overcoming the drawbacks of the continuous forcing approach, as the introduction of the force term at the discretization stage leads to a more stable and efficient algorithm (\citet{Mittaletal2005}). This method first introduced by \citet{mohdyusof1997}, has been developed in numerous original research works (see for examples \citet{Fadlun200035, Kim2001132, Balaras2004375, tairacolonius2007}) including a dedicated solver in OpenFOAM  (\citet{jasak14}). The drawback of these methods is that they are sensitive to the discretization, especially that of the time derivative. In this context, the semi-implicit treatment of the viscous terms to reduce the viscous stability constraint has a direct influence on the computation of the force term (\citet{Fadlun200035,Kim2001132}). \citet{Kim2001132} suggested to perform a first step explicitly to compute the force, and then to add the obtained force term to the equations, treated in a semi-implicit way. Although the method is computationally efficient, the velocity field and the force term are not evaluated at the same time instant in the algorithm, which can lead to stability issues. Another important aspect which is targeted in the present work is the analysis of moving boundaries. The related velocity fields generally suffer from spurious oscillations occurring during the time-marching of the algorithm, when a mesh element occupied by the flow suddenly becomes a \textit{solid} cell. In order to overcome these difficulties, \citet{uhlmann2005} proposed a direct forcing method combining the strengths of both continuous and direct forcing approaches. The method relies on the evaluation of the force term in the Lagrangian space, thus using  the delta functions originally proposed by Peskin. It has been successively improved by \citet{Pinelli20109073}, who introduced a new efficient quadrature for the spreading step and extended the method to non-uniform and curvilinear meshes. Owing to its modularity, stability, computational efficiency and accuracy in the analysis of moving/deformable configurations, this method has been identified as the best candidate to be implemented in the \textit{OpenFOAM} solver. Compared to the IBM method recently implemented in \textit{OpenFOAM} by \citet{jasak14}, the present approach appears to be more accurate and more  versatile for the study of unsteady/deforming structures, as it relies only on the accuracy of the interpolation and spreading steps, which are independent of the complexity of the geometry. 

Although it is not systematically mentioned explicitly in the literature, the application of direct forcing approaches in the context of incompressible flow solvers with predictor-corrector schemes is not straightforward. In fact, it is a two-constraints problem: on the one hand, the force term needed to impose the no-slip condition at the solid boundary must be calculated, and on the other hand, a divergence-free velocity at the boundaries must be satisfied. This means that enforcing divergence free conditions on the velocity affects the accuracy of the immersed boundary force at the wall. Although this issue has been claimed to be negligible by \citet{Fadlun200035}, it  may actually lead to significant differences depending on the configuration considered. It has been shown to systematically introduce a first-order error in time on the actual boundary values (\citet{domenichini2008}). A solution has been proposed by \citet{kajishimaetal2007} which changes the matrix structure of the Poisson problem solved to compute the value of the projector term (i.e., pressure or pressure correction), by directly imposing Neumann type conditions on the immersed boundary on the corresponding matrix terms. In order to avoid changing the matrix structure, \citet{tairacolonius2007} have suggested to use Lagrangian multipliers associated to boundary values to impose the expected velocity condition on the immersed boundary. Those Lagrangian multipliers are obtained solving a system derived from an algebraic splitting of the full spatial operator of the Navier-Stokes equations. In the present work, we choose an iterative scheme based on sub-iterations between (IBM) and pressure correction. This allows to use fast optimized Poisson solvers while  keeping control of the error made on both the velocity at the immersed boundary and the divergence of the velocity field.

The objectives of the present work are threefolds: (i) implement the improved IBM of \citet{Pinelli20109073} into \textit{OpenFOAM}, (ii) verify the numerical efficiency and accuracy of the new solver, (iii) validate the solver on well-documented test-cases of the literature. The paper is structured as follows: the numerical method is presented in section \ref{sec::Num_model} together with the geometry and governing equations. The implementation of the IBM in \textit{OpenFOAM } is detailed in section \ref{sec::DetailsIBM-OpenFOAM}, including the incorporation of the IBM in the PISO algorithm.  Preliminary verification work based on the analysis of the convergence of various errors is carried out in section \ref{sec::verification}. In section \ref{sec::validation} the solver is validated comparing flow simulations around fixed and moving circular cylinders to available data of the literature. The numerical performances are discussed in section and \ref{sec::IBM_scalability}, and concluding remarks and perspectives are provided in section \ref{sec::conclusions}.

\section{Numerical model}
\label{sec::Num_model}
\subsection{Geometrical model}
\label{sec::Config}
\begin{figure}[H]
\begin{center}
\includegraphics[width=0.7\linewidth]{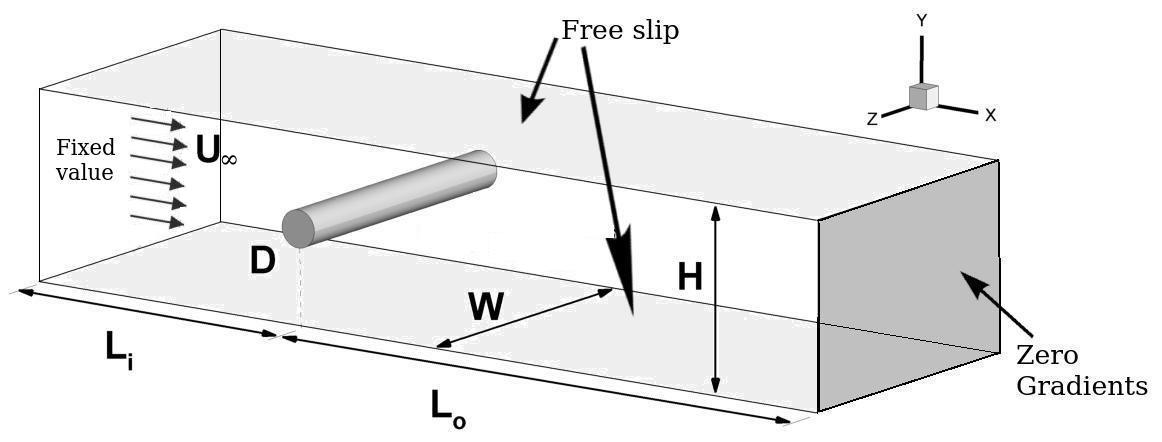}
\end{center}
\caption{  Flow configuration and computational domain \label{fig:FlowConfig}}
\end{figure}

Without any loss of generality, the paper focuses on the three-dimensional (3D) flow past a circular cylinder of diameter $D$, as shown in Figure \ref{fig:FlowConfig}.  The computational domain is a parallelipedic volume of height $H$, width $W$ and length $L_i+L_0$.
At the inlet, a steady uniform velocity is imposed along the streamwise direction $x$ together with a zero pressure gradient. A mass conservation condition is imposed at the outlet. We assume periodic conditions in the spanwise direction z. Free-slip boundary conditions on the velocity are applied at the top and bottom of the domain.

The incompressible flow of a viscous, Newtonian fluid is described by the dimensionless Navier Stokes equations written in a Cartesian frame of reference $(x, y, z)$:

\begin{equation}
\label{eq::NavierStokes}
\frac{\partial \textbf{u}}{\partial t} + \nabla \cdot (\textbf{uu})=-\nabla p+\dfrac{1}{Re} \nabla^2 \textbf{u}
\end{equation}
\begin{equation}
\label{eq::ConsMass}
\nabla  \cdot \textbf{u} = 0,
\end{equation}
where $\textbf{u}$ is the velocity vector, $p$ is a normalized pressure by the fluid density $\rho$ and $Re=\textbf{u}_{\infty} D/\nu$ is the Reynolds number with $\nu$ the kinematic velocity. The incompressibilty of the fluid is guaranteed by the continuity equation (\ref{eq::ConsMass}). Taking the divergence of equation (\ref{eq::NavierStokes}) and using the continuity equation (\ref{eq::ConsMass}) yields classically a Poisson equation for the pressure :

\begin{equation}
\label{eq::Poisson}
\nabla^2 p = - \nabla \cdot (\textbf{u}\nabla \textbf{u})
\end{equation}
Equations (\ref{eq::NavierStokes}) and (\ref{eq::Poisson}) couple pressure and velocity in an elliptic manner that requires specific numerical algorithms.

\subsection{\textit{OpenFOAM}}
\label{sec::OpenFOAM}

The present work relies on classical \textit{OpenFOAM} setup for space and time discretization :

\begin{itemize}

\item The governing equations are discretized in space using finite volume (FV) on fixed and structured meshes, composed of hexaedral elements (\citet{jasak96}). The time discretization is based on the first-order implicit Euler scheme which has been chosen for its simplicity to simulate low-Reynolds numbers flows.\footnote{Note, the IBM implementation does not depend on the time discretization scheme,  meaning that all numerical details provided herein carry over to more accurate second-order \textit{OpenFOAM} schemes, such as backward Euler or Crank-Nicolson.}  A constant time-step approach has been chosen, with time step values ensuring the stability of the algorithm algorithm, a point further discussed below.

\item The velocity-pressure coupling is solved by the built-in solver \textit{pisoFoam} in which the stabilizing extra term on the mass flux in the pressure correction loop is set to zero (see details in the recent paper by \citet{Vuorinen14}). The solver is thus the classical Pressure Implicit with Splitting of Operators (PISO) algorithm described in the paper by  \citet{Ferziger96}. Three and one iterations were set for a PISO loop and for non orthogonal corrections (\citet{Devilliers06}) respectively.

\item Linear algebraic systems are solved using the Diagonal Incomplete LU Preconditioned Biconjugate Gradient \textit{DILUPBG} (for the momentum equation (\ref{eq::NavierStokes})) and the Diagonal Incomplete Cholesky Preconditioned Conjugate Gradient \textit{DICPCG} (for the Poisson equation (\ref{eq::Poisson}) ). For the present simulations involving low Reynolds numbers and regular structured meshes, no preconditionning has been used. For all independent variables, the required accuracy is $10^{-7}$ at each time step.

\end{itemize}

\subsection{The immersed boundary method (IBM)}
\label{sec::IBM}

We recall that the IBM formulation chosen in this work is the discrete forcing approach of \citet{Pinelli20109073}. The Navier--Stokes equations are discretized on a fixed mesh (Eulerian) while the solid boundary is discretized by a set of Lagrangian markers free to move over the Eulerian mesh, depending on the motion of the solid.
As in traditional direct forcing methods, the target velocity $U^d$ is directly imposed at the boundary nodes. This velocity is equal to either the local fluid velocity or zero depending on wether the solid moves or is at rest.  

\subsubsection{Calculation of the body force term on the Lagrangian markers: the interpolation step}
The body force is computed in the Lagrangian space, i.e. at all Lagrangian markers. On the $s^{\text{th}}$ Lagrangian marker and at $(n+1)^{th}$ time step, the force term, $\mathbf{F}_s^{n+1}$, is given by:

\begin{equation}
\label{eq:force1}
\mathbf{F}_s^{n+1} = \frac{\mathbf{U}_s^d-\mathcal{I}[\mathbf{u}^*]_s }{\Delta t}
\end{equation}
where $\mathbf{U}_s^d$ is the target velocity on the $s^{\text{th}}$ Lagrangian marker.  $\mathcal{I}[\mathbf{u}^*]_s$ stands for the interpolation on the $s^{\text{th}}$ Lagrangian marker of the fluid velocity known on the Eulerian mesh at $n^{th}$ time step, and computed without any force term. Therefore, $\mathbf{u}^*$ is a predictive velocity computed advancing in time the momentum equation (\ref{eq::NavierStokes}) without any boundary. As presented in \citet{lietal2015}, the discrete expression of the interpolation operator is given by :
\begin{equation}
\label{eq:interp}
\mathcal{I}[\mathbf{u}^n]_s = \sum_{j\in D_s} \mathbf{u}_{j}^n \delta_h(\mathbf{x}_{j} - \mathbf{X}_s) \Delta v
\end{equation}
where the $j$-index refers to the discrete value of the fluid velocity on the Eulerian mesh, $\mathbf{X}_s$ refers to the coordinates of the $s^{\text{th}}$ Lagrangian marker and $\Delta v$ refers formally to an Eulerian quadrature, i.e. $\Delta v=\Delta x \Delta y \Delta z $ for the case of a Cartesian uniform mesh. 
The interpolation kernel $\delta_h$ is the discretized delta function used in \citet{romaetal1999} :
\begin{equation}
{\delta}_h(r)  \left\{
\begin{aligned}
&\frac{1}{3}\left(1 + \sqrt{-3r^2 + 1} \right) \ &0 \leq r \leq 0.5\\
&\frac{1}{6}\left[5 - 3r - \sqrt{-3(1 - r)^2 + 1}\right] \ &0.5 \leq r \leq 1.5\\
&0 \ &\textnormal{otherwise}
\end{aligned}
\right.
\label{eq:kernel_Roma}
\end{equation}
It is centered on each Lagrangian marker $s$ and takes non-zero values inside a finite domain $D_s$, called the support of the  $s^{\text{th}}$ Lagrangian marker.
\subsubsection{Calculation of the body force term on the Eulerian mesh: the spreading step}
Once the force term is computed from equation (\ref{eq:force1}), one needs to transfer its value to the Eulerian mesh. This is done by the spreading step, which is the inverse operation of the interpolation. The value of the force term evaluated on the Eulerian mesh, $\mathbf{f}^{n+1}(\mathbf{x}_j)$, is given by:
\begin{equation}
\label{eq:force2}
\mathcal{S}[\mathbf{F}_k^{n+1}] = \mathbf{f}^{n+1}(\mathbf{x}_j) = \sum_{k \in D_j} \mathbf{F}_k^{n+1} \delta_h(\mathbf{x}_{j} - \mathbf{X}_k) \boldsymbol{\epsilon}_k
\end{equation}
The $k$-index refers to a loop over the Lagrangian markers whose support contains the Eulerian node $j$. $\boldsymbol{\epsilon}_k$ is the Lagrangian quadrature, which is calculated solving a linear system :
\begin{equation}
\label{eq:forceA}
A \boldsymbol{\epsilon} = \mathbf{1}
\end{equation}
 where the vectors $\boldsymbol{\epsilon}=(\epsilon_1, \dots, \epsilon_{N_s} )^T$ and $\mathbf{1}=(1, \dots, 1)^T$ have a dimension of $N_s$, $N_s$ being the number of Lagrangian markers, and $A$ is the matrix defined by the product between the $k^{\text{th}}$ and the $l^{\text{th}}$ interpolation kernels such that:
\begin{equation}
A_{kl} = \sum_{j\in D_l} \delta_h(\mathbf{x}_j-\mathbf{X}_k) \delta_h(\mathbf{x}_j-\mathbf{X}_l)  
\end{equation}
The last step consists in solving again the momentum equation (\ref{eq::NavierStokes}) by adding the force term $\mathbf{f}^{n+1}$ computed using equation (\ref{eq:force2}) at the right hand side of the equations. For further details, see the papers by \citet{Pinelli20109073,lietal2015}.

\section{The (IBM) implementation in \textit{OpenFOAM}}
\label{sec::DetailsIBM-OpenFOAM}

The (IBM) presented in section \ref{sec::IBM} is implemented in \textit{pisoFOAM}. This code version will be named (IBM)-\textit{OpenFOAM} in the following. The code implementation is discussed and illustrated from the flow solution past a circular cylinder at Re = 30.

One of the of the most problematic issues regarding the implementation of the IBM in predictor-corrector codes is that the velocity field at the immersed boundaries must satisfy both the no-slip and the divergence-free condition. In order to resolve this two-constraint problem, we use the following procedure at each time step $n$:
\begin{enumerate}

\item{The momentum Navier--Stokes equations in the predictor step are solved without force term. A first estimate of the velocity $\widehat{\textbf{u}}$ is thus obtained from:
	\begin{equation}
	\label{eq::NavierStokes_predictor}
	\frac{\partial \widehat{\textbf{u}}}{\partial t} + \nabla \cdot (\widehat{\textbf{u}}\widehat{\textbf{u}})=-\nabla p+\dfrac{1}{Re} \nabla^2 \widehat{\textbf{u}}
	\end{equation}
	
}
\item{The force $\mathbf{F}_s$ is calculated from the velocity field $\widehat{\textbf{u}}$ using equation (\ref{eq:force1}). The force $\mathbf{f}$ acting on the Eulerian mesh is then derived by equation (\ref{eq:force2}).
}
\item{The momentum equation is solved a second time, including the immersed boundary force term:
\begin{equation}
	\label{eq::NavierStokes_guess_poisson}
	\frac{\partial \textbf{u}^{\star,1}}{\partial t} + \nabla \cdot (\textbf{u}^{\star,1}\textbf{u}^{\star,1})=-\nabla p+\dfrac{1}{Re} \nabla^2 \textbf{u}^{\star,1}-f(\widehat{\textbf{u}})
	\end{equation}
where $\textbf{u}^{\star,1}$ is the final velocity obtained in the predictor step, which now accounts for the IBM force.
}
\item{The pressure field is solved in the corrector step as a solution to the Poisson equation :
\begin{equation}
\label{eq::PoissonPISO_loop}
\nabla^2 p^{\star,1} = - \nabla \cdot (\textbf{u}^{\star,1} \nabla \textbf{u}^{\star,1}) + \nabla \cdot f(\widehat{\textbf{u}})
\end{equation}
}

\item{  The predicted mass fluxes and velocity field are updated from the obtained pressure field, i.e.:
 
\begin{equation}
\label{eq::UpdatePISO}
\textbf{u}^{\star,2} = g \left(\textbf{u}^{\star,1}, \, \nabla p^{\star,1} , \, f(\widehat{\textbf{u}}) \right)
\end{equation}
where $g$ is a function tied to the strategy used for the representation of the non-linear term of the Navier--Stokes equations all related details being provided in \ref{subsec::PISO_IBM_implementation_details}.
}
\item{
The system is then advanced in time.
Steps 4 and 5 are iteratively repeated until the convergence criteria set by the user are fulfilled.

\begin{equation}
\label{eq::UFinal}
\textbf{u}^{n+1}=\textbf{u}^{\star,M}
\end{equation}
}
\end{enumerate}
 Moreover, a selective procedure for the Lagrangian markers has been implemented to ensure a well-conditioned system (equation (\ref{eq:forceA}) described in the paper by \citet{Pinelli20109073}). The values of $\boldsymbol{\epsilon}$ and the capability of the method on the selection of points can be checked by a specific flag which activates a diagnostic implemented in the code on the Lagrangian grid.

\begin{figure}[H]
\begin{center}
\includegraphics[width=0.7\linewidth]{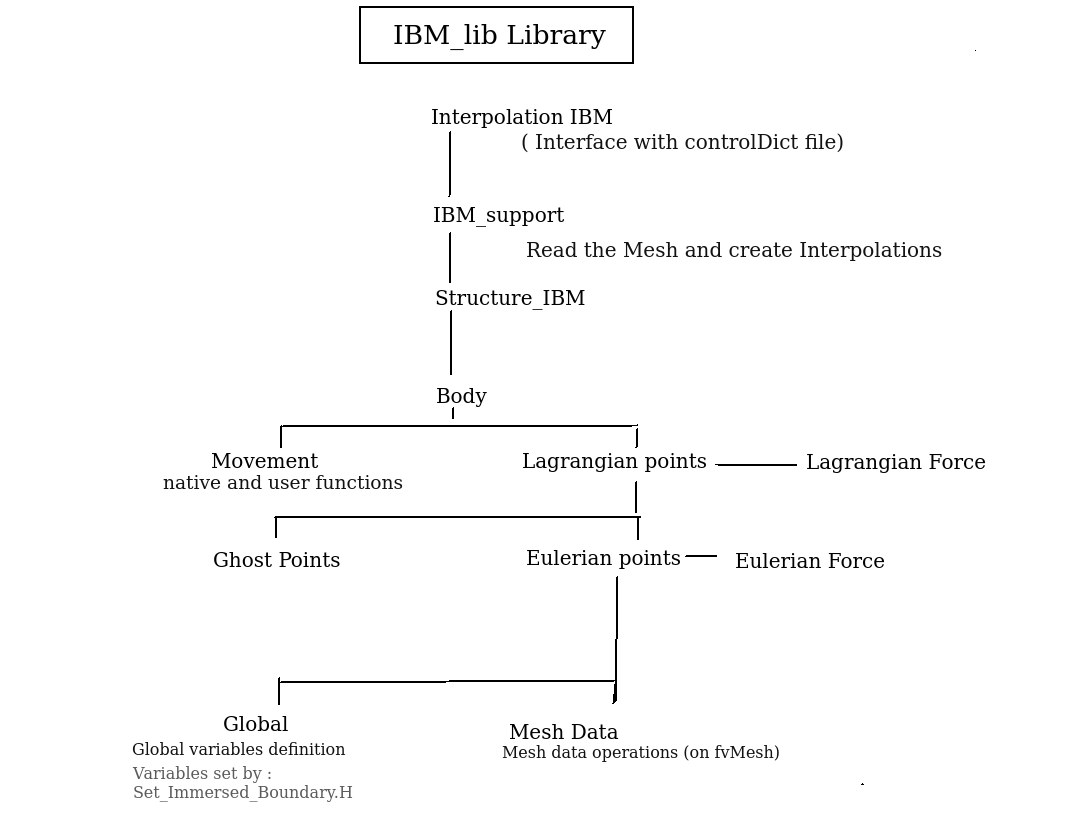}
\end{center}
\caption{ Tree of the IBM Library IBM\textunderscore lib for OpenFOAM.} \label{fig:IBM_lib}
\end{figure}

The part of the program that calculates the interpolation an the spreading is object oriented and can be included in any other solver by the user (programmed as a library). This means that the present development herein presented for the solver \textit{pisoFoam}, and summarized by the scheme in Figure \ref{fig:IBM_lib}, can be straightforwardly extended to any other solver based on a predictor-corrector algorithm. The scheme of the code is shown in Figure \ref{fig:IBM_lib}. The library has been succesfully tested in OpenFoam 2.1.1, 2.2.1, 2.2.2, 2.2.X and 3.0.1.

\subsection{\textit{pisoFOAM} amendments}
\label{subsec::PISO_modification}

\begin{figure}[H]
\begin{center}
\begin{tabular}{cc}
\includegraphics[width=0.4\linewidth]{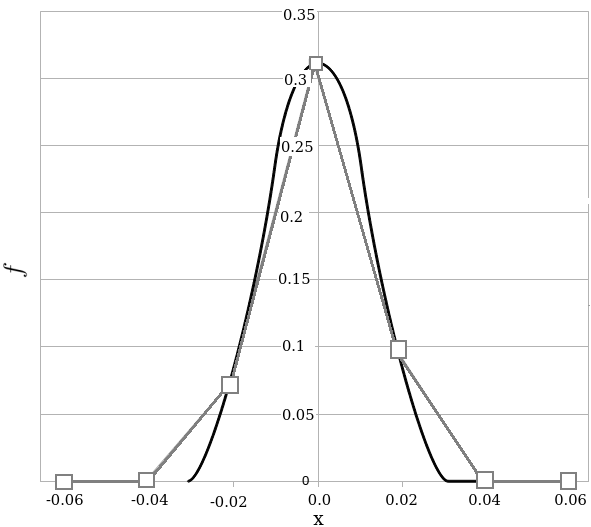} &
\includegraphics[width=0.5\linewidth]{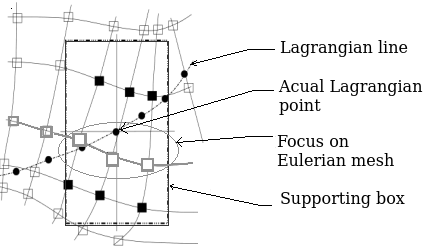}
\end{tabular}
\end{center}
\caption{Example of Eulerian discretization of the IBM force term $f$ :  comparison between 2$^{nd}$ order discretization using a centered scheme and the kernel analytical solution (left). Zoom on the mesh discretization (right). Flow simulation past a fixed cylinder  at $Re = 30$.} \label{fig:force_discretization}
\end{figure}

\begin{figure}[H]
\begin{center}
\includegraphics[width=0.5\linewidth]{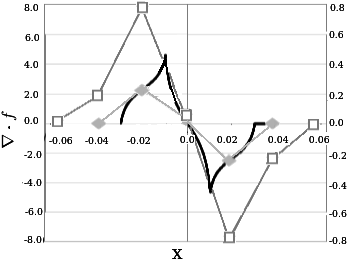}
\end{center}
\caption{Derivative of the force term using three different kinds of interpolation. The results of a 2$^{nd}$ order discretization using a centered scheme are shown in empty grey symbols and refer to the left scale 10 times higher than the other results. The full grey symbols represent the new derivative calculated with the kernel function. The black line represents the derivative theoretical value for the flow simulation past a fixed cylinder  at $Re = 30$. \label{fig:force_discretization_1}}
\end{figure}

The correct calculation of both the velocity divergence around the structure and the flux requires an appropriate integration of the body force term into the momentum equation. The discretization of the structure boundary leads to non-negligible errors due to the sharpness of the body force term (ideally discretized over 3 markers). This is illustrated  in Figures \ref{fig:force_discretization} and \ref{fig:force_discretization_1}, which show the interpolation of the IBM force term and how it impacts its derivative at one given Lagrangian marker.

\begin{figure}[H]
\begin{center}
\includegraphics[width=1\linewidth]{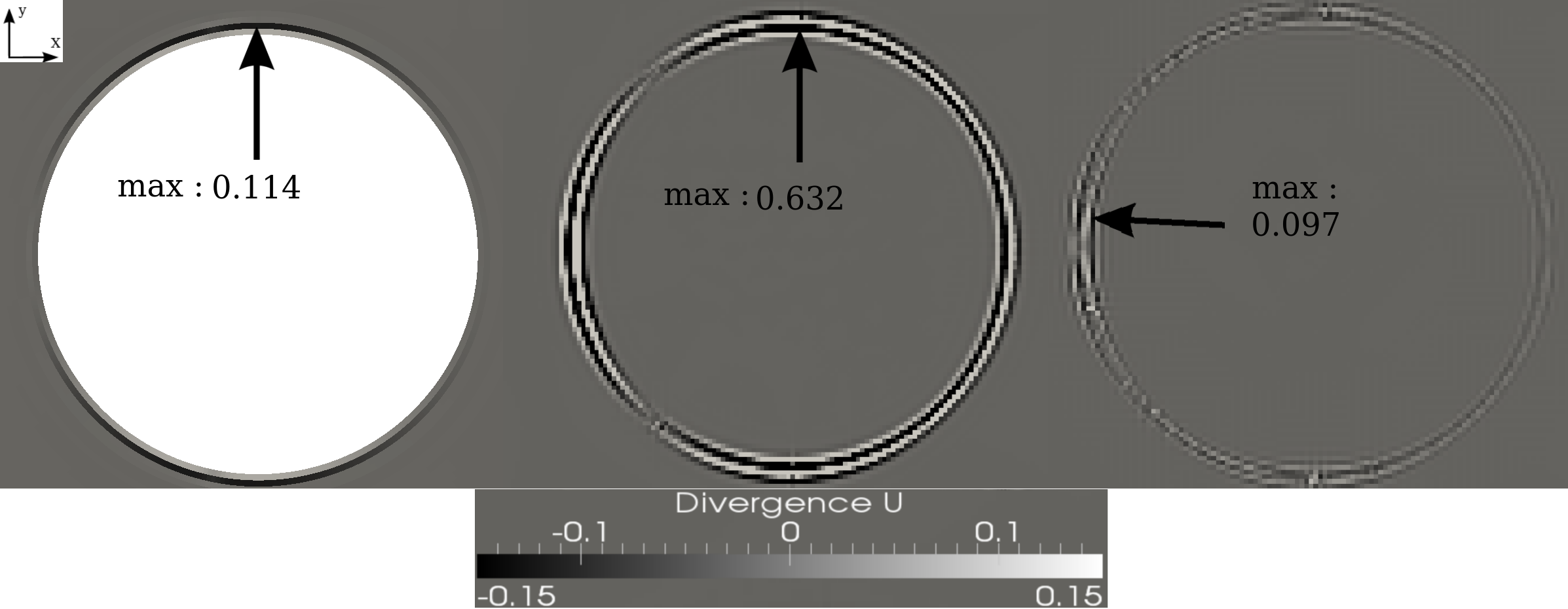}
\end{center}
\caption{ Plots of the velocity divergence around a 2D cylinder at Re = 30. Classical no-slip boundary condition (left). IBM calculations with 312 Lagrangian markers: without (center) and with (right) correction of the derivative. Corresponding meshes are those of Figure \ref{fig:div_u_2}. \label{fig:div_u_1}}
\end{figure}

Two solutions can be considered to overcome this issue. First, the stencil can be enlarged using additional points to derivate and interpolate the force term. However, this solution leads to a more  diffuse (and thus, less accurate) definition of the boundary. Instead, in this work the computation of the divergence of the momentum equation (in the PISO loop) and the interpolation of the fluxes is achieved by an hybrid calculation involving an analytical resolution (with the kernel function equation (\ref{eq:kernel_Roma})) of the quantities involving the force term (singular quantities). Comparative results  in Figure \ref{fig:div_u_1} show that the maximum of the error on the divergence is reduced by a factor of almost $85$\%  (compare to a body fitted mesh) when using the correction on the derivative. The presented divergences are calculated at the end of the PISO loop with the function \textit{fvc::div} on the velocity field.

\begin{figure}[H]
\begin{center}
\includegraphics[width=0.7\linewidth]{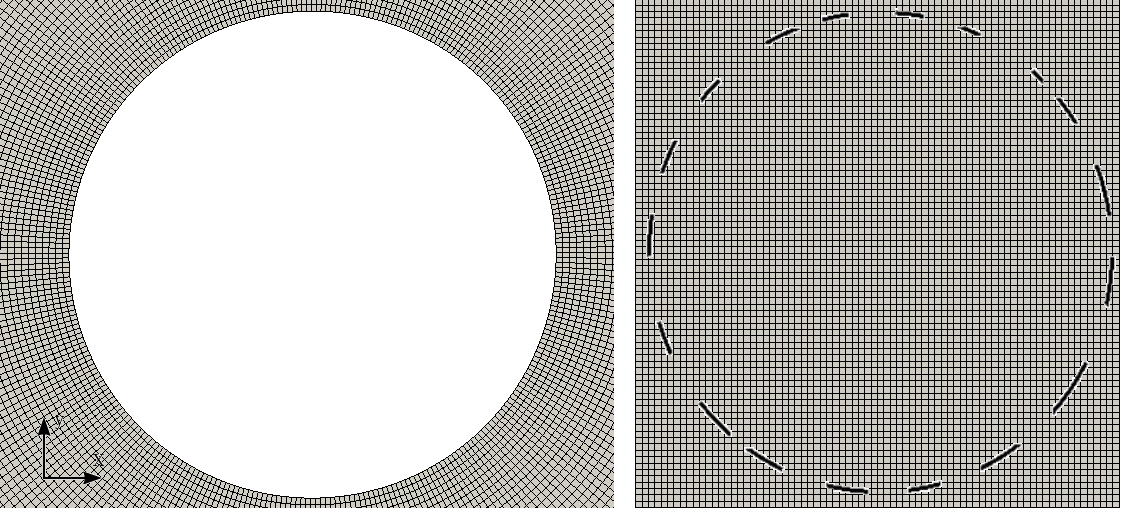}

\end{center}

\caption{ Zoom on the different meshes used for the estimates of the divergence. Body fitted mesh around the cylinder when using no-slip boundary condition. $\Delta$x = 1.10$^{-2}$ around the cylinder (left). Cartesian mesh with IBM (right).\label{fig:div_u_2}}

\end{figure}

\section{Solver verification}
\label{sec::verification}

The code verification has been performed in 2D using both the method of the manufactured solution for the velocity and the pressure, and a grid convergence study for the flow solution past a circular cylinder at $Re=30$.

\subsection{Manufactured solution}
Polynomial functions  $f(x,y)$, $g(x,y)$ and $h(x,y)$ have been chosen for the two components of the velocity $\textbf{u}_a(f(x,y);g(x,y))$ (divergence free), and the pressure $p_a(h(x,y))$, respectively such that:

\begin{eqnarray}
\label{eq::manufactured_solution_f}
 \mathbf{u_a} = \left\lbrace
 \begin{array}{l*1{c}}
f(x,y)  =  (1-0.01x^2)^2(1-0.03y^2)(1-0.01y^2)\\ \ \ \ \ \ \ \  \ \ \ \ \  -0.02(1-0.01x^2)^2(y-0.01)(y-0.01y^3)  \\
 g(x,y)  =  0.5+0.04x(1-0.01x^2)(y-0.01y^3)(1-0.01y^2)
\end{array}\right.
\end{eqnarray}
\begin{equation}
p_a = h(x,y)  =  f(x,y) g(x,y)
\end{equation}

Three errors have been defined to verify different steps in the solver:

\begin{enumerate}
\item $e_{F_{IBM}}$ is the error on the estimate of the (IBM) force term defined by  equation (\ref{eq:force1}) (during Step 2 of the IBM/PISO solver described in Section \ref{sec::DetailsIBM-OpenFOAM}) and integrated on the body, hence computed as:

\begin{equation}
\label{eq::e1}
e_{F_{IBM}}= \mid \sum_{k \in D_j} \mathbf{(F_k - F_a)}\boldsymbol{\epsilon}_k \mid
\end{equation}

where :

\begin{eqnarray}
\mathbf{F}_{a} &=& \frac{\mathbf{U}_k^d-\mathbf{U_a}}{\Delta t}
\end{eqnarray}

and $\mathbf{U_a}$ is the value of the analytical solution $\mathbf{u_a}$, defined below by equation (\ref{eq::manufactured_solution_f}), evaluated on the Lagrangian markers. \\

\item $e_{noslip}$ is the error on the estimate of the no-slip condition at the boundary of the obstacle. This error is evaluated during the calculation of the IBM force term on the Eulerian mesh (end of Step 2 of the IBM/PISO solver).
It is defined as the L$_{\infty}$ norm of the difference between the  velocity on one Lagrangian marker (equation (\ref{eq:interp})), and the Eulerian velocity that has been spread and re-interpolated, i.e. : 
 
\begin{equation}
\label{eq::e2}
e_{noslip}=\parallel U_s-\mathcal{I}[\mathcal{S}[U_k]]_s \parallel _{\infty}
\end{equation}

\item $e_{u_{tot}}$ is the error on the velocity at the end of the PISO loop (Step 6 of the IBM/PISO solver Section \ref{sec::DetailsIBM-OpenFOAM}). It is calculated in terms of  both the $L_2$ and $L_{\infty}$ norms: 
\begin{equation}
\label{eq::e3_1}
e_{u_{tot}/L_2 } = \parallel u  - u_a \parallel _2
\end{equation}
\begin{equation}
\label{eq::e3_2}
e_{u_{tot}/L_{\infty} } = \parallel u  - u_a \parallel _{\infty}
\end{equation}
\end{enumerate}

The verification is made in five steps summarized below: 
\begin{itemize}
\item Computation of $\textbf{u}$ and $p$ according to:

\begin{eqnarray}
\label{eq::manufactured_source_term}
\frac{\partial \textbf{u}}{\partial t} + \nabla \cdot (\textbf{uu}) = \nabla p + \dfrac{1}{Re} \nabla^2 \textbf{u}+ S_a  \\
S_a=\frac{\partial \mathbf{u_{a}}}{\partial t} + \nabla \cdot (\mathbf{u_au_a})- \nabla p_a - \dfrac{1}{Re} \nabla^2 \mathbf{u_a}
\end{eqnarray}

\item Computation on the Lagrangian markers of the analytical values of the IBM force term $F_a$ using $\textbf{u}_a$ and of the IBM force term $F_s$ using the interpolated velocity $u$ from the former step.

\item Calculation of $e_{F_{IBM}}$ and  $e_{noslip}$ using equations  (\ref{eq::e1}) and  (\ref{eq::e2}).
\item Spreading of the residual force $F_s-F_a$ on the Eulerian grid.
\item Execution of steps 3 to 6 of the PISO algorithm (section \ref{sec::DetailsIBM-OpenFOAM}) and calculation of $e_{u_{tot}/L_2 }$ and $e_{u_{tot}/L_{\infty} }$.
\end{itemize}

The errors $e_{F_{IBM}}$, $e_{noslip}$ and $e_{u_{tot}}$ have been calculated for two 2D flows past a circular and a square cylinder (to quantify the impact of the geometry on the method) of diameter and side $L/5$, respectively, $L$ being the size of the computational domain. For a $(10 \times 10)$-domain, four  uniform grids have been tested, corresponding to $\Delta x = \Delta y = 5 \times10^{-2}$, $3.3\times10^{-2}$, $2.5\times10^{-2}$ and $1.25\times10^{-2}$. 
To properly evaluate the interpolation error, the square cylinder is not aligned with the cells centers. Thus,
Lagrangian markers have been chosen to be located between two Eulerian nodes. 

Results are shown on Figure \ref{fig:Square_cyl_conv}. All  errors decrease when the mesh is refined. The error $e_{noslip}$ exhibits a second-order rate of convergence whereas $e_{u_{tot}/L_{2} } $ and $e_{u_{tot}/L_{\infty} } $ only exhibit a rate of convergence between 1 and 2 for both geometries. Finally, $e_{F_{IBM}}$ exhibits a rate of convergence that depends on the geometry (as could have been expected), namely 1 for the square cylinder and nearly 2 for the circular cylinder. \\

\begin{minipage}[b]{0.545\linewidth} 

\begin{center}
\includegraphics[width=1\linewidth]{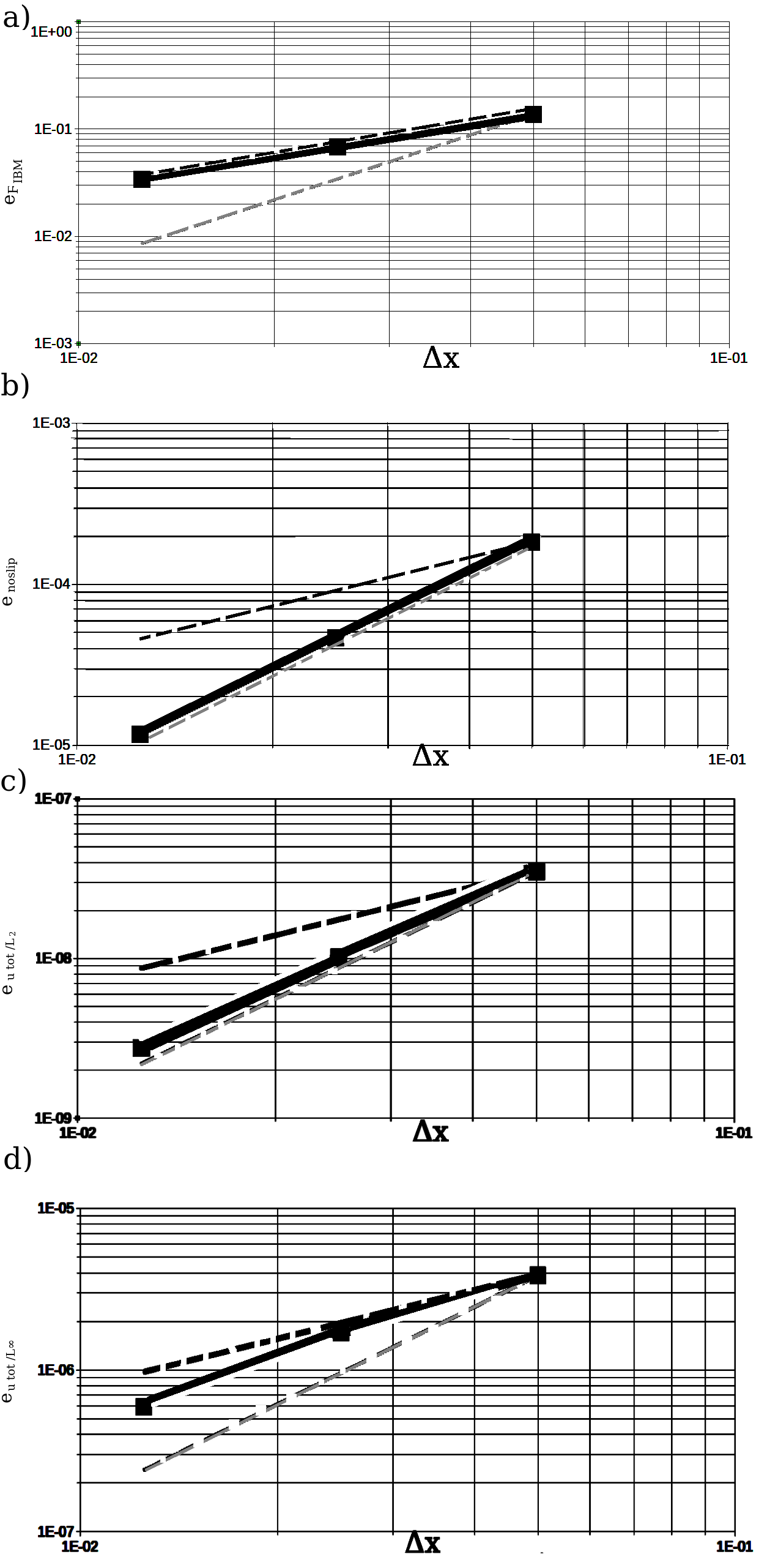}
\
\end{center}

  \end{minipage}
\begin{minipage}[b]{0.55\linewidth} 

\begin{center}
\includegraphics[width=1\linewidth]{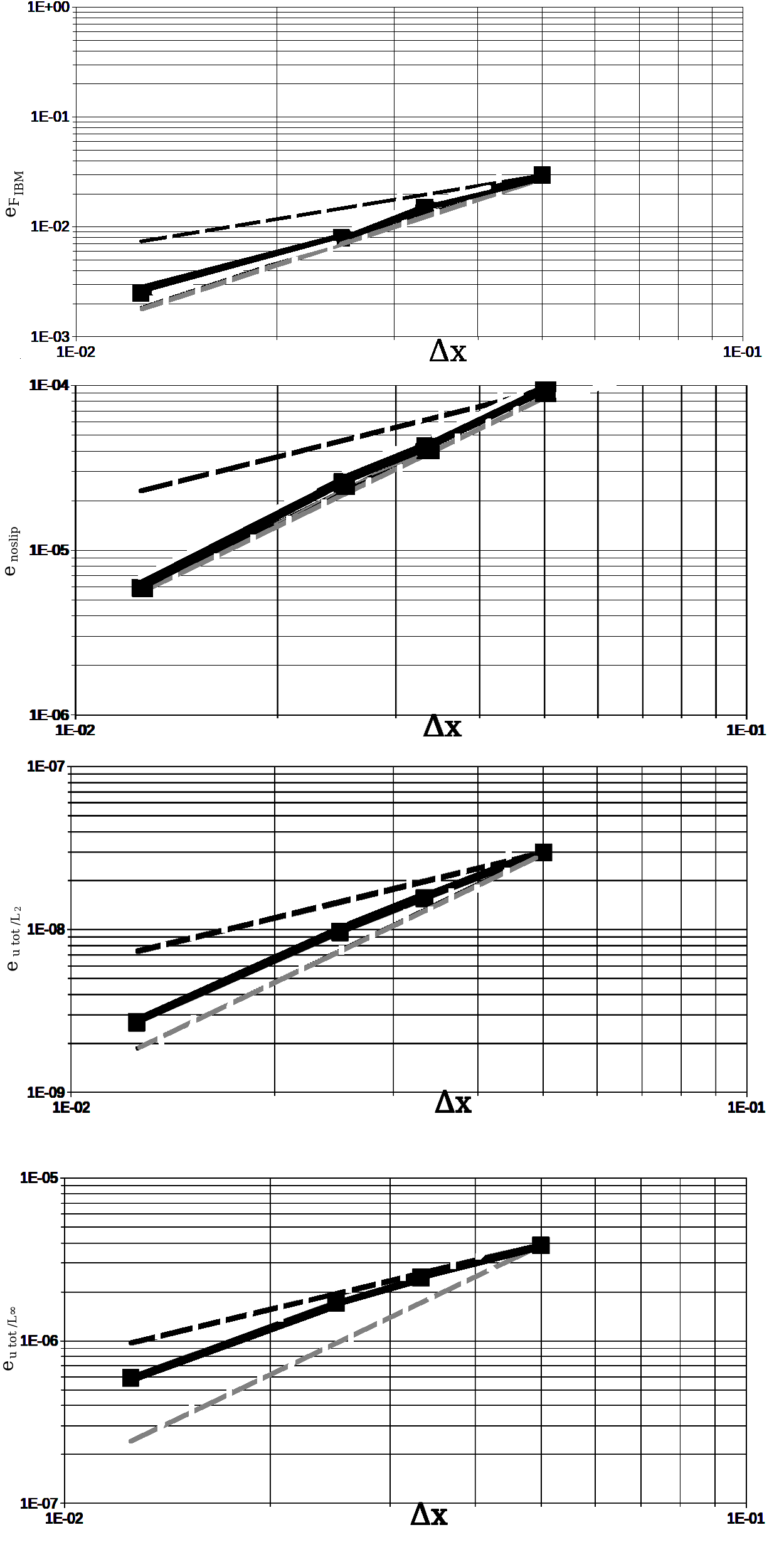}
\end{center}

  \end{minipage}

\begin{figure}[H]

\caption{ Log-log plots of $e_{F_{IBM}}$ , $e_{noslip}$and $e_{u_{tot} } $ (full line) as a function of the mesh refinement, for flows past a square  (left) and a circular (right) cylinder: $e_{F_{IBM}}$ (a), $e_{noslip}$ (b),  $e_{u_{tot}/L_{2} } $ (c) and $e_{u_{tot}/L_{\infty} } $ (d). The dashed black and grey lines show the slopes of order 1 and 2, respectively.   \label{fig:Square_cyl_conv} }
\end{figure}

\subsection{Grid convergence on the flow past a cylinder at $Re=30$}
\begin{figure}[H]
\begin{center}
\includegraphics[width=0.65\linewidth]{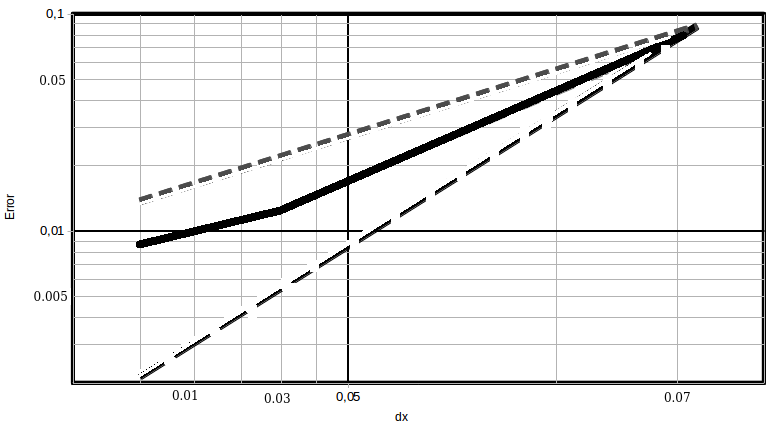}
\end{center}
\caption{ Log-log plot of the error on the drag coefficient computed for various  mesh refinements comparing the value of the drag coefficient $C_D$ to the reference value computed with the finest mesh. The dashed black and grey lines show the slopes of order 1 and 2, respectively.  \label{fig:convergence_cd}}
\end{figure}

Four grids have been used  corresponding to $\Delta x = \Delta y = 8 \times10^{-2}$, $4\times10^{-2}$, $2\times10^{-2}$ and $1\times10^{-2}$. The solution computed on the finest mesh is the reference solution. The error is estimated from the drag coefficient $C_D$ defined in equation (\ref{eq::coeffs}) and with respect to its value on the finest mesh, see Figure \ref{fig:convergence_cd}. The error descreases when the mesh is refined, with an order between 1 and 2.

\section{Solver validation}
\label{sec::validation}

The solver validation is performed using two- and three-dimensional (2D/3D) simulations of flows past a circular cylinder of diameter $D$ and at various Reynolds numbers $(Re=U_{\infty} D/\nu)$ corresponding to well-documented test cases of the literature, summarized in Tables \ref{tab:re30} to \ref{tab:fixed_cylinder3D}. The Strouhal number, drag and lift coefficients are defined by:
\begin{eqnarray}
\label{eq::coeffs}
C_D = \frac{2F_d}{\rho u_{\infty}^2 D}\; \; , \; \;  C_L = \frac{2F_l}{\rho u_{\infty}^2 D} \; \; , \; \;  St=\dfrac{Df_v}{u_{\infty} },
\end{eqnarray}
where $f_v$ is the shedding frequency and $F_d$ and $F_l$ are the drag force and lift force per unit length, respectively, computed by integrating the immersed boundary force term in the Lagrangian space.

\subsection{Computational details} 

\begin{figure}[H]
\begin{center}
\includegraphics[width=1\linewidth]{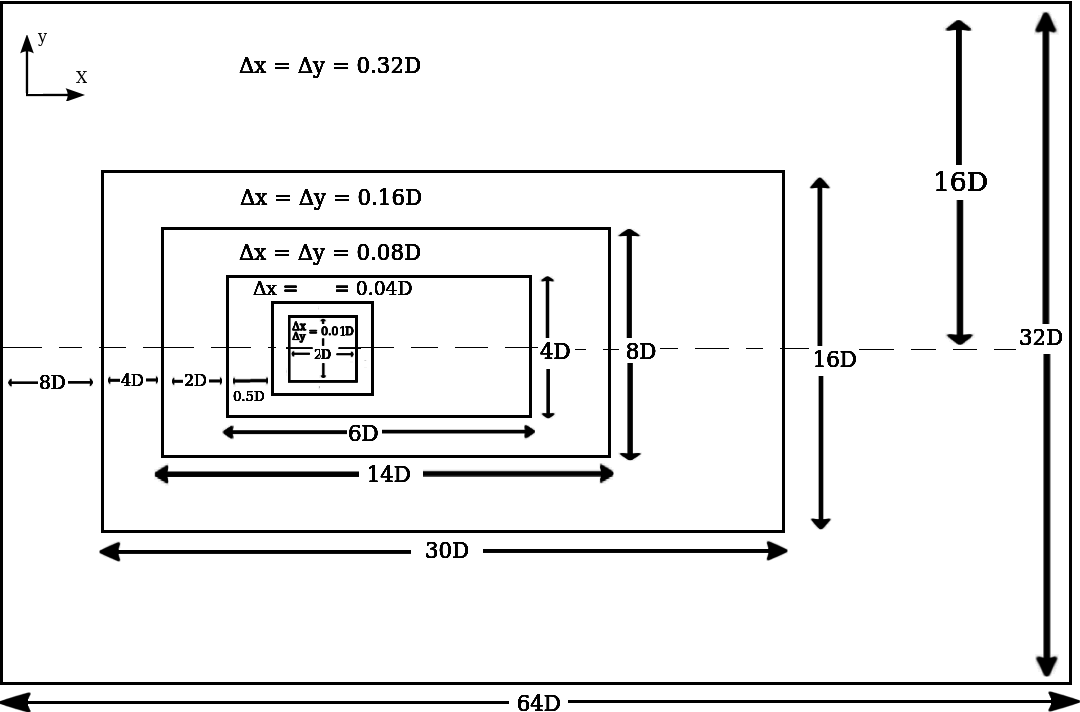}
\\ \ \\  \ \\
\includegraphics[width=1\linewidth]{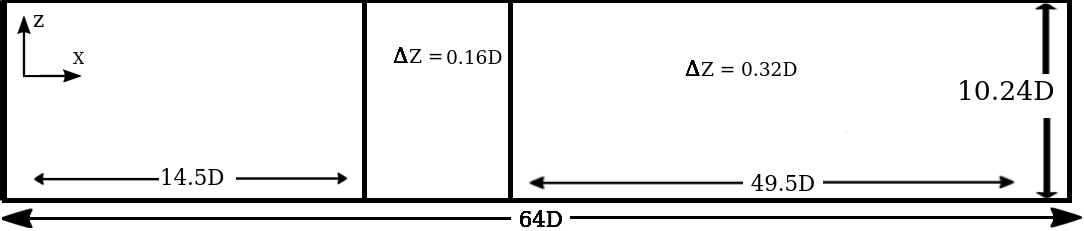}

\end{center}
\caption{\label{fig:computational_domain1} Computational domain decomposition and grid spacings. $(x,y)$-plane (top) and spanwise direction (bottom).}

\end{figure}

The center of the cylinder is the origin of the domain at $(0,0)$. The dimensions of the computational domain are those proposed by \citet{Pinelli20109073} and \citet{Vanella09},
namely  $[-16D, 48D]\times[-16D, 16D]\times[-5.12D, 5.12D]$ in the streamwise $(x)$, vertical $(y)$ and spanwise $(z)$ directions (Figure \ref{fig:computational_domain1}).

The grid is uniform  in the neighborhood of the cylinder, i.e. in the region $-D \leq x \leq D$ and $-D \leq y \leq D$. For 3D computations, the 2D mesh has been extruded in the spanwize direction. Details pertaining to the resolution, as well as the number of Lagrangian markers and their relative spacing with respect to the Eulerian mesh are given in Table \ref{tab:simulation_case}. Outside this region, the mesh size is stretched with a factor of $2.0$ on five grid levels in the $(x,y)$-plane (as shown in Figure  \ref{fig:computational_domain1}).

\begin{table}[!htbp]
\caption{\label{tab:simulation_case} Mesh resolutions in the neighborhood of the cylinder: 2D cases 1 and 2 $[-D, D]\times[-D, D]$, 3D case 3 $[-D, D]\times[-D, D]\times[-5.12D, 5.12D]$. The $\alpha$ parameter defines the ratio of the distance between Lagrangian markers over the local Eulerian grid size \citet{Pinelli20109073}.}
\begin{tabular}{cccc}
\toprule
Case & Resolution & Lagrangian markers & $\alpha$ \\ 

\hline

1 & $\Delta x=\Delta y=0.02D$ &  147  & 1.061\\
2 & $\Delta x=\Delta y=0.01D$ &  312  & 1.004 \\
3 & $\Delta x=\Delta y=0.02D, \Delta z=0.16D$ &  9792  & 1.004 \\
\bottomrule
\end{tabular}
\end{table}

All 2D and 3D simulations have been performed on 12 and 96 cpu, respectively. The CFL has been fixed to $0.5$ and the number of PISO loop to 3. Simulations time varies from 24 hours (2D simulations) to 168 hours (3D simulations) depending on the mesh size and the flow regime.

%
%
%

\begin{figure}[!htbp]
    \setcaptionwidth{5.0
in}
\begin{minipage}[b]{1.0\linewidth} 
    \centering 
\includegraphics[width=3.2 in]{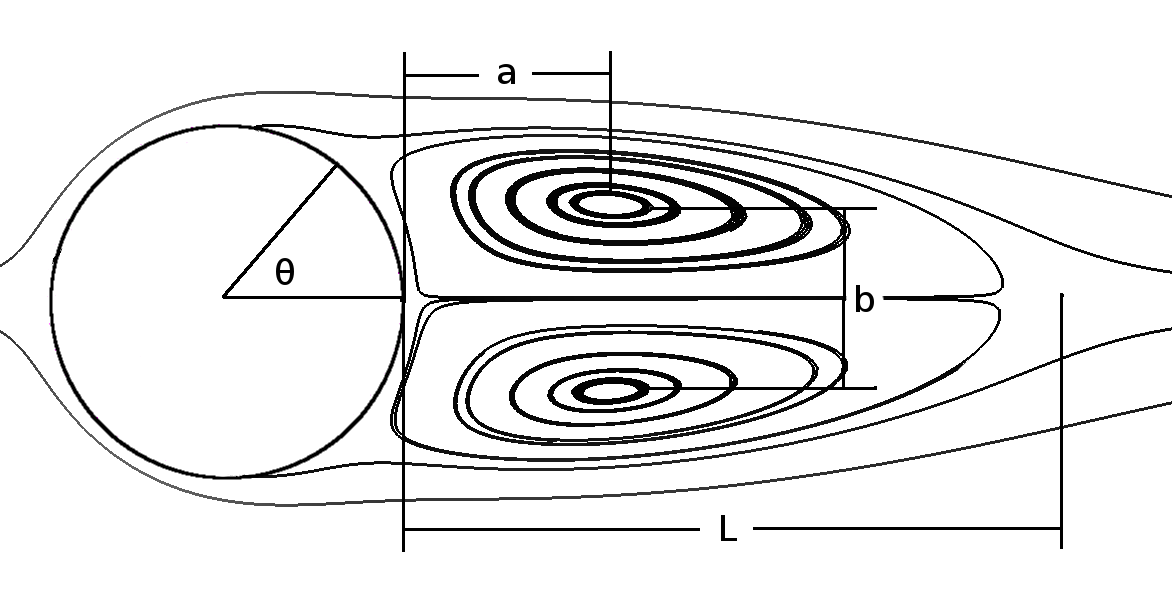} 
  \end{minipage}
\caption{\label{Re30_wake} Characteristic geometrical parameters in the steady regime. $L$ is the length of the recirculation, $a$ is the distance between the cylinder and the recirculations centers, $b$ is the vertical distance between the two recirculation centers, and $\theta$ is the separation angle measured from the rear stagnation point.} 
\end{figure}

\subsection{2D Flow around a fixed cylinder} 
Three 2D simulations have been performed at Reynolds numbers, $Re=30$, $100$, and $185$.
\subsubsection{ Steady flow - Re=30} 

At $Re=30$, the flow is characterized by a steady recirculating region located just behind the cylinder. All characteristic geometrical parameters defined on Figure \ref{Re30_wake} compare well with the data of the literature reported in  Table \ref{tab:re30}, with differences less than 6\% for the most  refined grid.

\begin{table}[H]
\setcaptionwidth{6.5in}
  \caption{\label{tab:re30} Geometrical parameters of the wake and drag coefficient for the configuration of a fixed cylinder at $Re=30$. Numerical and experimental data from literature are provided for comparison.}
\hspace*{-15mm}
  \begin{tabular}{c|c|ccccc}
    \toprule
    \multicolumn{2}{c|}{}&L/D&a/D&b/D&$\theta^{o}$&$C_D$\\ \hline
    \multirow{2}{*}{\textbf{Present ($Re=30$)}}&$\Delta x=\Delta y=0.02D$&\textbf{1.66}&\textbf{0.556}&\textbf{0.53}&\textbf{47.80}&\textbf{1.78}\\
    \cline{2-2}
                         &$\Delta x=\Delta y=0.01D$&\textbf{1.64}&\textbf{0.55}&\textbf{0.53}&\textbf{48.40}&\textbf{1.77}\\
    \hline
    \multicolumn{2}{c|}{\citet{Pinelli20109073}}&1.70&0.56&0.52&48.05&1.80\\
    \multicolumn{2}{c|}{\citet{Blackburn99}}&-&-&-&-&1.74\\
    \multicolumn{2}{c|}{\citet{coutanceau77}}&1.55&0.54&0.54&50.00&-\\
    \multicolumn{2}{c|}{\citet{Tritton59}}&-&-&-&-&1.74\\
    \bottomrule
  \end{tabular}
\end{table}
\subsubsection{ Unsteady flow - Re=100-185} 

\begin{figure}[H]
    \setcaptionwidth{6.2in}
\begin{minipage}[b]{1.0\linewidth} 
\centering
\includegraphics[width=2.5in]{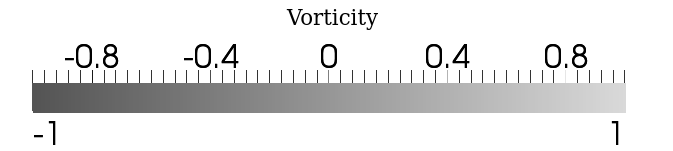} 
  \end{minipage}\\
\hspace{-13cm}\small{(a)}\\
\begin{minipage}[b]{1.0\linewidth} 
\centering
\includegraphics[width=3.4in]{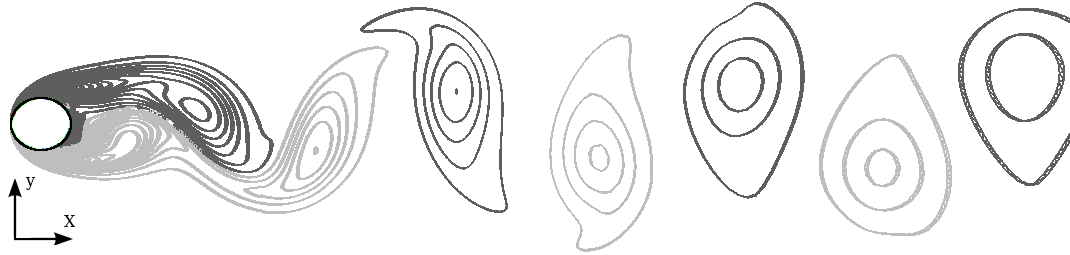} 
  \end{minipage}\\
\hspace{-13cm}\small{(b)}\\
\begin{minipage}[b]{1.0\linewidth} 
\centering
\includegraphics[width=3.4in]{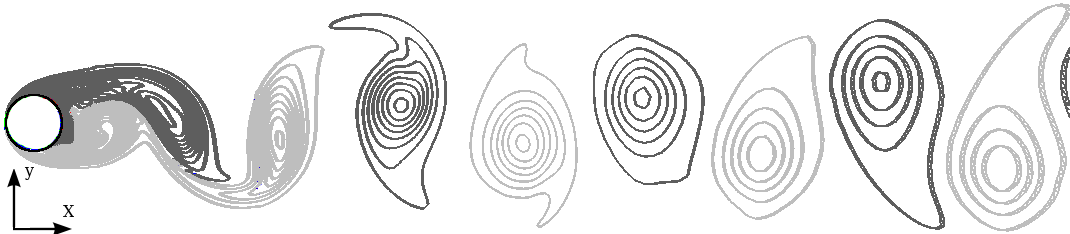} 
  \end{minipage}
\caption{\label{Vorticity_Re100-185} Vorticity countours evidencing the shedding of large-scale vortices in 2D flow past a fixed circular cylinder at $Re=100$ (a) and $Re=185$ (b).}
\end{figure}

\begin{figure}[H]
    \setcaptionwidth{5.5in}
\includegraphics[width=1.0\linewidth]{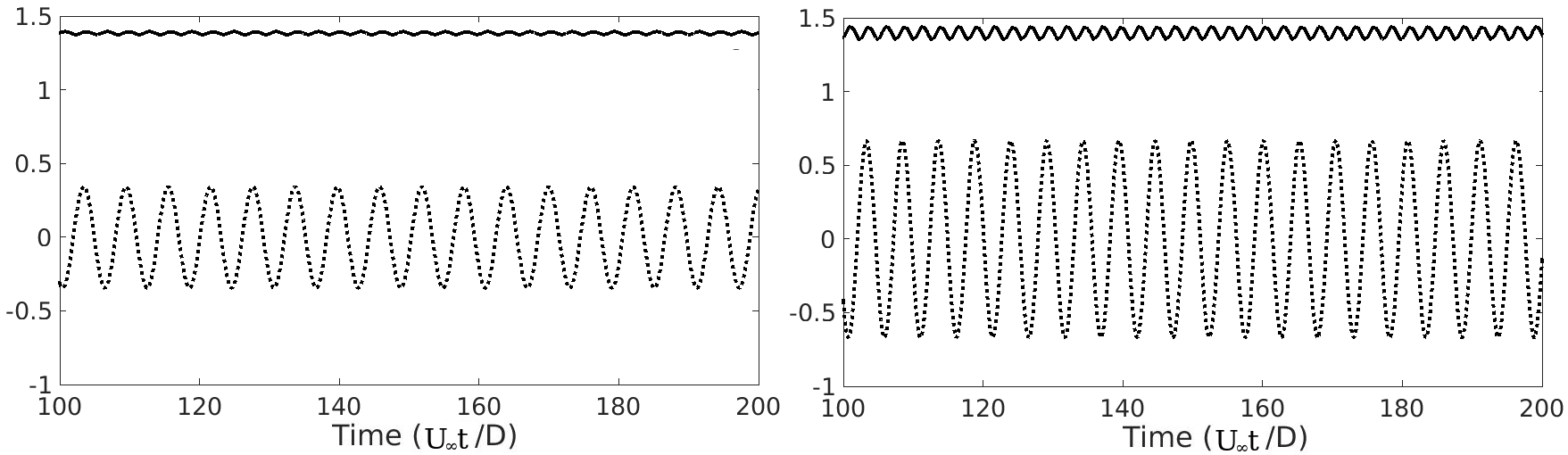} 
\caption{\label{fig:cd_cl} Temporal evolution of $C_D$ (full line) and $C_L$ (dashed line) for the 2D flow at $Re=100$ (left); $Re=185$ (right).}
\end{figure}

\begin{figure}[H]
    \setcaptionwidth{7.0in}
\begin{minipage}[b]{1\linewidth} 
\centering
\includegraphics[width=0.8\linewidth]{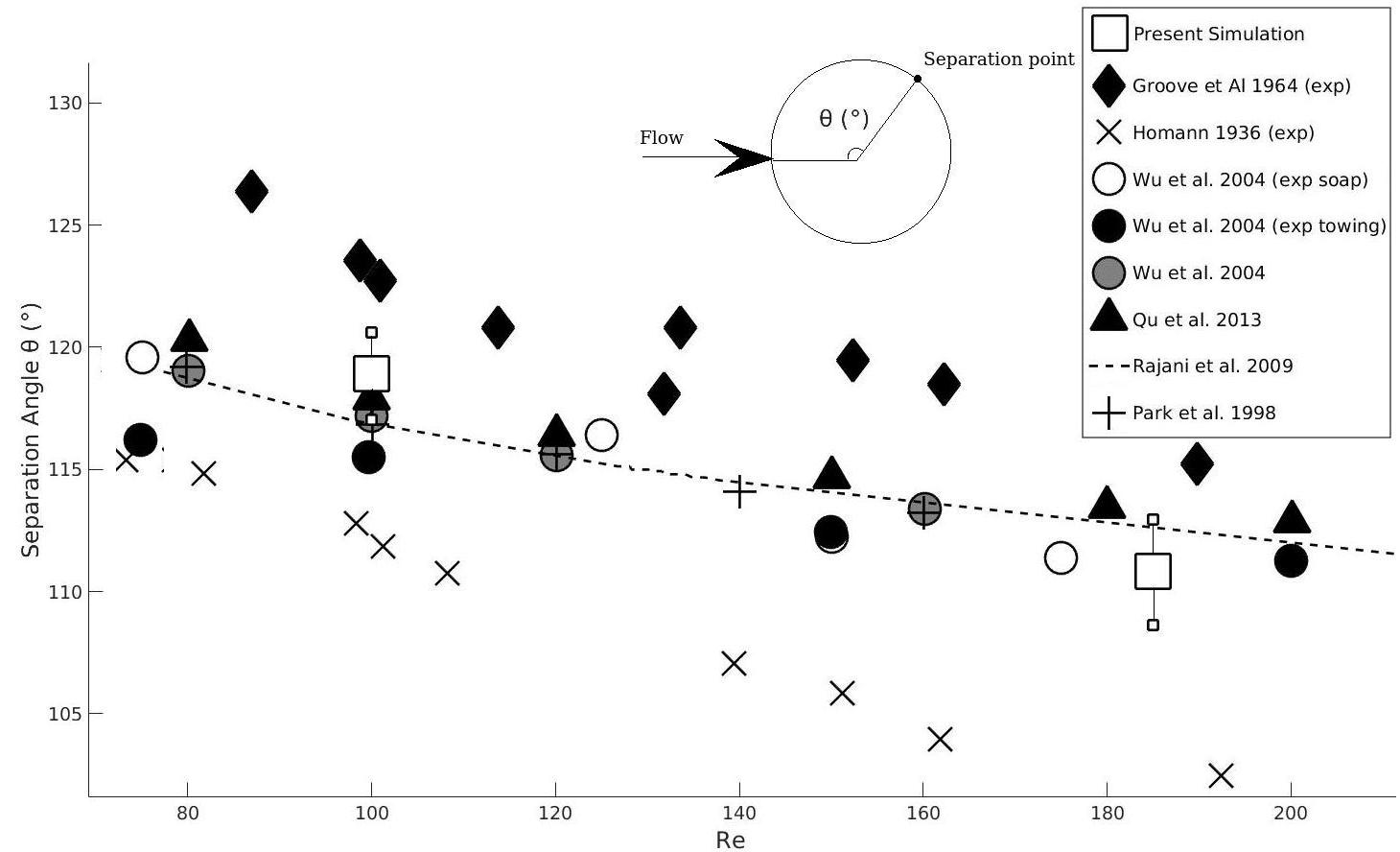} 
  \end{minipage}

\caption{\label{fig:separation} Mean separation angle as a function of the Reynolds number. Error bars corresponding to the min/max values achieved during the duration of the averaging process are shown by the vertical lines on either side of the square symbols.}
\end{figure}

Simulations in 2D unsteady regimes with vortex shedding have been performed at $Re=100$ and $185$, i.e. above $Re_c=40$ for the transition to unsteadiness according to \citet{Williamson88} and \citet{Norberg94}. The vorticity contours shown in Figure \ref{Vorticity_Re100-185} exhibit the well-known Karman vortex street featuring the periodic shedding of vortices, convected and diffused away from the cylinder. The topology of the solutions compares well with that reported in  several reference studies, see for instance the papers by \citet{Guilmineau02,Pinelli20109073}. The corresponding time evolutions of $C_D$ and $C_L$ are plotted in Figure \ref{fig:cd_cl}  and show that the amplitude of the lift and drag fluctuations increase with the Reynolds number, in good agreement with the paper by \citet{Guilmineau02}.  For both Reynolds numbers, the Strouhal number, the mean drag (computed over 10 time periods) and the rms lift coefficients compare well with the literature data summarized in Table \ref{tab:fixed_cylinder_re100_185}.  Figure \ref{fig:separation} puts further emphasis on the mean separation angle, which is also well predicted by the present simulations.

\begin{table}[H]
\setcaptionwidth{6.5in}
  \caption{\label{tab:fixed_cylinder_re100_185} Mean drag, rms lift coefficients and Strouhal number for 2D flow past a fixed cylinder at  $Re=100$ and $Re=185$. Numerical and experimental data from the literature are provided for comparison.}
\hspace*{-15mm}
  \begin{tabular}{c|c|ccc}
       \toprule
    \multicolumn{2}{c|}{}&$C_D$&$C_L^{rms}$&$S_t$\\ \hline
    \multirow{2}{*}{\textbf{Present ($Re=100$)}}&$\Delta x=\Delta y=0.02D$&\textbf{1.38}&-&\textbf{0.165}\\
    \cline{2-2}
                         &$\Delta x=\Delta y=0.01D$&\textbf{1.37}&-&\textbf{0.165}\\
    \hline
    \multicolumn{2}{c|}{\citet{Blackburn99}}&1.35&-&-\\
    \multicolumn{2}{c|}{\citet{Barkley96}}&-&-&0.165\\
    \multicolumn{2}{c|}{\citet{Williamson96}}&-&-&0.164\\
    \multicolumn{2}{c|}{\citet{Henderson95}}&1.35&-&-\\
    \multicolumn{2}{c|}{\citet{Norberg94}}&-&-&0.164\\
\midrule
    \multirow{2}{*}{\textbf{Present ($Re=185$)}}&$\Delta x=\Delta y=0.02D$&\textbf{1.387}&\textbf{0.436}&\textbf{0.198}\\
    \cline{2-2}
                         &$\Delta x=\Delta y=0.01D$&\textbf{1.379}&\textbf{0.427}&\textbf{0.198}\\
    \hline
    \multirow{2}{*}{\citet{Pinelli20109073}}&$\Delta x=\Delta y=0.005D$&1.430&0.423&0.196\\
    \cline{2-2}
                         &$\Delta x=\Delta y=0.01D$&1.509&0.428&0.199\\
    \cline{1-2}
    \multicolumn{2}{c|}{\citet{Vanella09}}&1.377&0.461&-\\
    \multicolumn{2}{c|}{\citet{Guilmineau02}}&1.287&0.443&0.195\\
    \multicolumn{2}{c|}{\citet{Lu96}}&1.310&0.422&0.195\\
    \multicolumn{2}{c|}{\citet{Williamson88}}&-&-&0.193\\
    \bottomrule
  \end{tabular}
\end{table}

\subsection{Flow around a 3D cylinder} 

\begin{figure}[H]
\begin{center}

\includegraphics[width=0.47\linewidth]{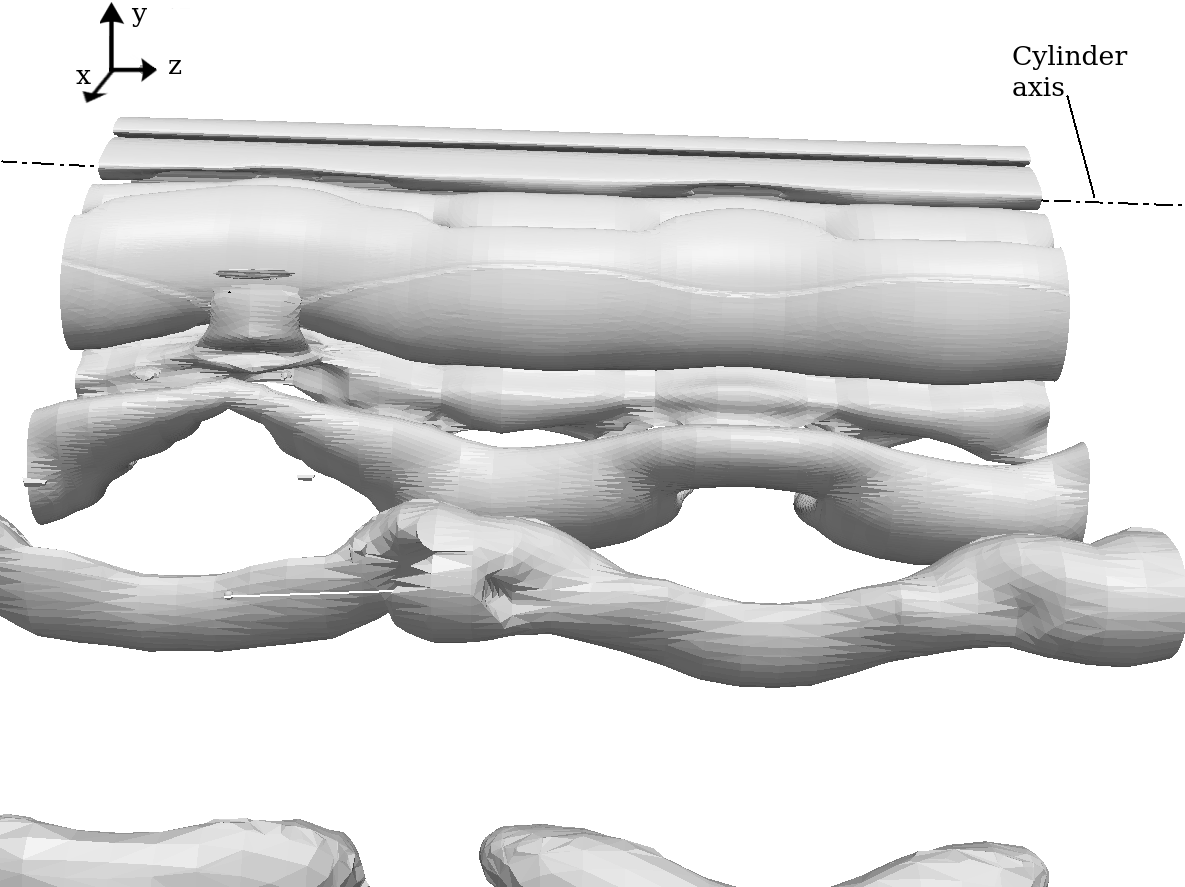}
\includegraphics[width=0.47\linewidth]{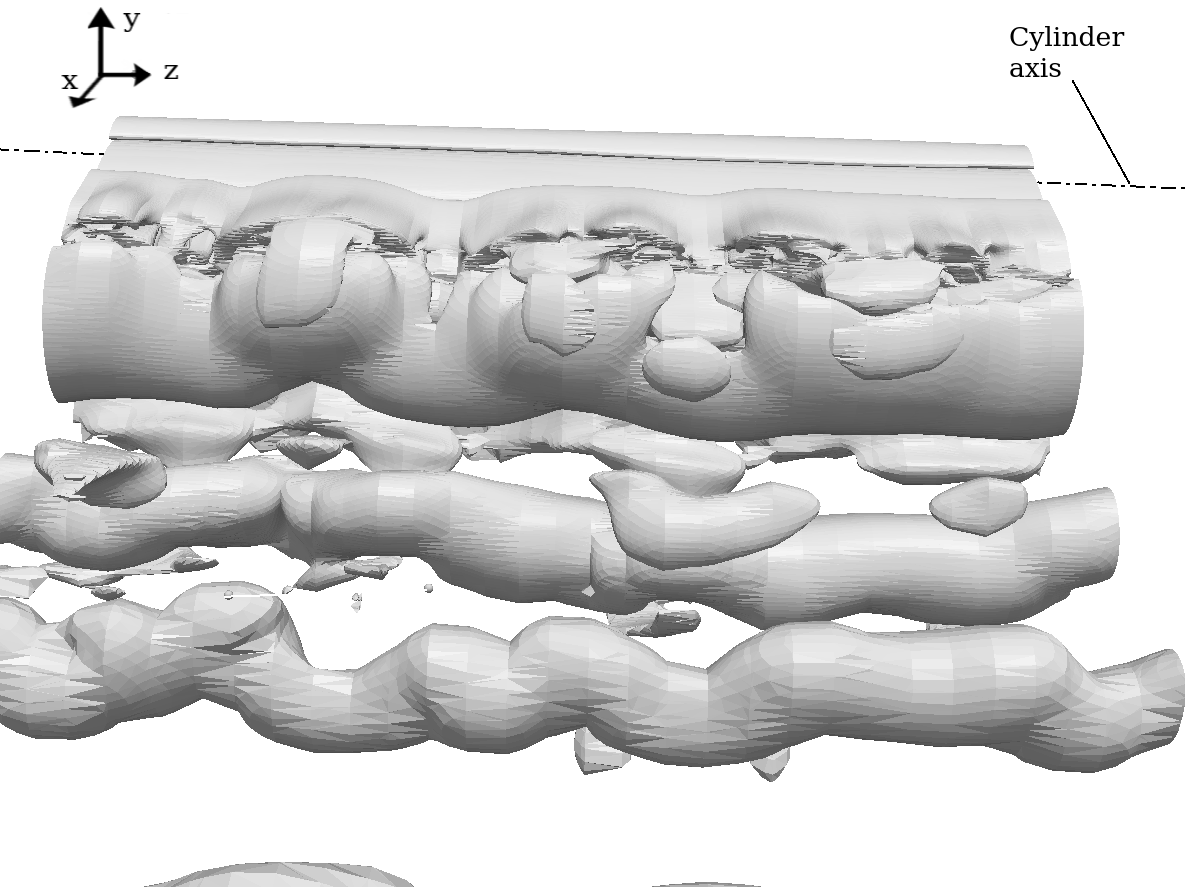}
\end{center}
\caption{ \label{fig:lambda_det} Iso-surfaces of the instantaneous Q-criterion $(-0.8 < Q < 0.8)$  at $Re = 200$ (left) and $Re = 300$ (right).}
\end{figure}

\begin{figure}[H]
    \setcaptionwidth{5.5in}
\includegraphics[width=1.0\linewidth]{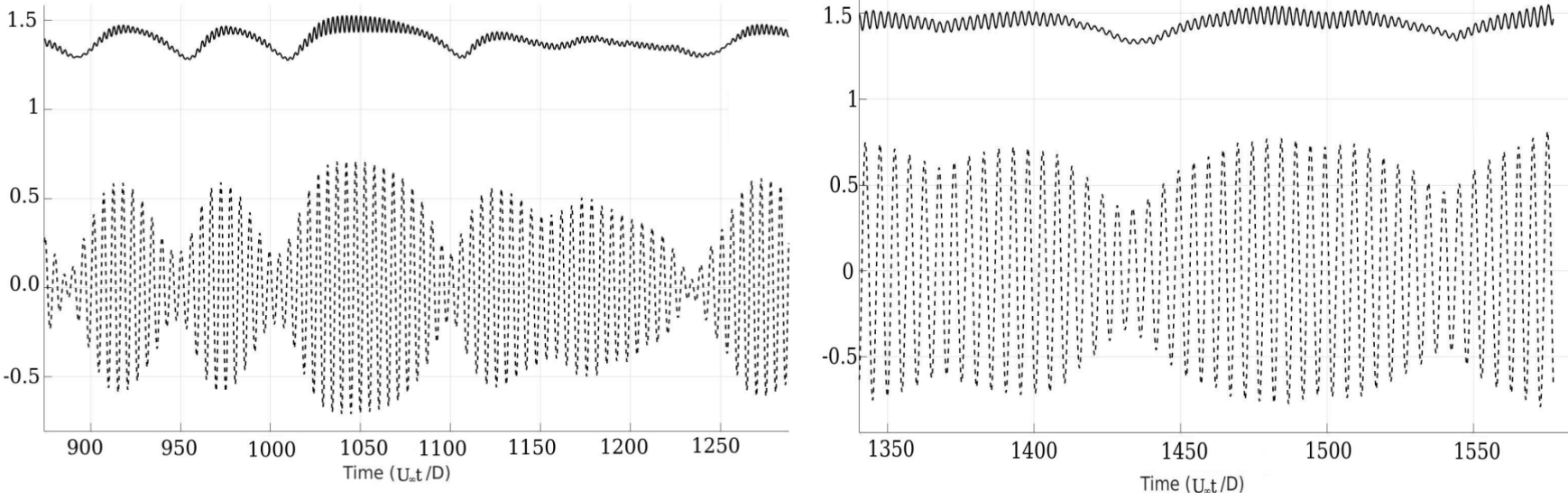} 
\caption{\label{fig:cd_cl3d} Temporal evolution of $C_D$ (full line) and $C_L$ (dashed line) for the 3D flow at $Re = 200$ (left); $Re = 300$ (right).}
\end{figure}

In order to show the capacity of the code to accurately predict 3D unsteady flows, additional simulations have been performed at $Re=200$ and $300$, i.e., above the critical value $Re_c=190$  for the transition to 3D flow, and within the range of Reynolds numbers where the 3D pattern transitions from mode A to mode B, according to the reference study of \citet{Williamson96}. The present simulations predict well the occurrence of 3D vortex shedding, as shown by the instantaneous $Q$-criterion iso-surfaces in Figure \ref{fig:lambda_det}. When increasing Reynolds number from $Re=200$ to $Re=300$, the solution shows a strong decrease of the spanwise wavelength $\lambda_z$, from $\lambda_z/D \simeq4.5$ to $\lambda_z/D \simeq1.25$ as previously observed by \citet{Williamson96} at the transition between mode A and mode B. The temporal evolution of $C_D$ and $C_L$ in Figures \ref{fig:cd_cl3d} show a modulated behaviour characteristic of these 3D flows, all values being in agreement with the literature data, as seen from Table \ref{tab:fixed_cylinder3D}.

\subsection{Flow around a oscillating cylinder} 

Finally, in order to assess the capacity of the solver to encompass moving obstacles, a 2D simulation of flow past a sinusoidally moving cylinder is performed at $Re=500$. Following \citet{Blackburn99}, the cylinder is forced to oscillate in the vertical direction at a fixed amplitude ratio of $A = 0.25$ and with a frequency ratio of $F (= f_o/f_v) = 0.975 $ (with $f_o$ the frequency of the forced oscillation). The computational domain is the same as for the fixed simulation, with $\Delta x=\Delta y=0.02D$ around the cylinder. The shedding frequency $f_v$ is obtained from a preliminary flow simulation past a fixed cylinder at $Re=500$.

\begin{center}

\begin{table}[H]
\setcaptionwidth{6.5in}
  \caption{\label{tab:fixed_cylinder3D}  Mean drag, rms lift coefficients and  Strouhal number for 3D flow past a fixed cylinder at $Re=200$ and $Re=300$. Numerical and experimental data from the literature are provided for comparison.}
\hspace*{-15mm}
\begin{tabular}{c|c|ccc}

    \toprule
    \multicolumn{2}{c|}{}&$C_D$&$C_L^{rms}$&$St$\\ \hline

    \multirow{1}{*}{\textbf{Re=200}}&$\Delta x=\Delta y=0.02D \ \ \& \  \ \Delta Z=0.16D$&\textbf{1.384}&\textbf{0.346}&\textbf{0.1802}\\
   
    \hline
    \multicolumn{2}{c|}{\citet{Rajani09}}&1.338&0.4216&0.1936\\
    \multicolumn{2}{c|}{\citet{Qu13}}&1.24&0.339&0.1801\\
    \multicolumn{2}{c|}{ \citet{Williamson96} (exp.)}&-&-&0.1800\\
    \multicolumn{2}{c|}{Pinelli (Intern Communication)}&1.371&0.163&0.1915\\

\midrule
    \multirow{1}{*}{\textbf{Re=300}}&$\Delta x=\Delta y=0.02D \ \ \& \  \ \Delta Z=0.16D$&\textbf{1.43}&\textbf{0.453}&\textbf{0.198}\\
    \hline
    \multicolumn{2}{c|}{\citet{Rajani09}}&1.28&0.499&0.195\\
    \multicolumn{2}{c|}{\citet{Mittal95}}&1.26&0.38&0.203\\
    \multicolumn{2}{c|}{ \citet{Williamson96}(exp.)}&-&-&0.203\\
    \multicolumn{2}{c|}{ \citet{Norberg93}(exp.)}&-&0.435&0.203\\
        \multicolumn{2}{c|}{ \citet{Wieselsberger22}(exp.)}&1.22&-&-\\
    \bottomrule
  \end{tabular}

\end{table}
\end{center}

\begin{figure}[H]
\begin{center}
\includegraphics[width=0.52\linewidth]{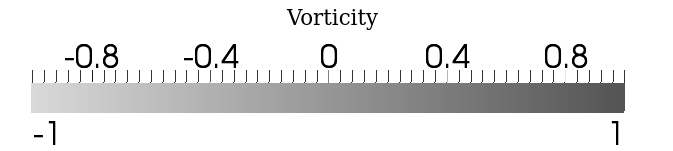}

\includegraphics[width=0.6 \linewidth]{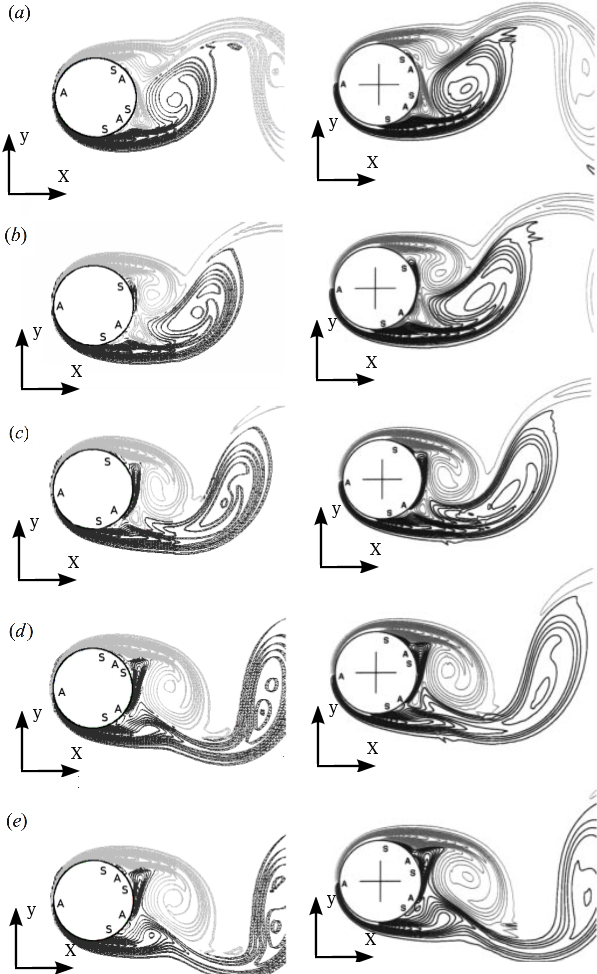}
\end{center}
\caption{\label{Re500_vorticity} Instantaneous contours of vorticity for the 2D flow past an oscillating cylinder at Re=500 : present results (left columns) vs. results obtained by \citet{Blackburn99} (right column) at five instants spreading over half of the shedding cycle.}

\end{figure}

\begin{figure}[H]
\begin{center}
\includegraphics[width=0.8\linewidth]{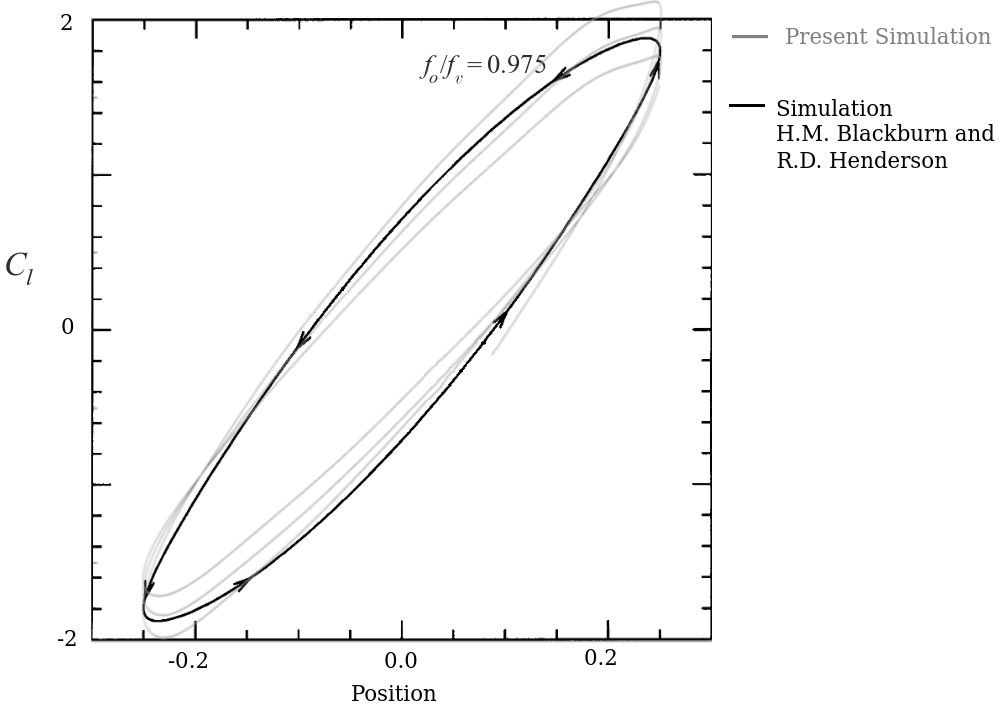}
\end{center}
\caption{ \label{Re500_displcmt_oscl} Lift coefficient $C_L$ as a function of the cylinder displacement for the 2D flow past an oscillating cylinder at Re=500 : present results (grey line) vs. results obtained by \citet{Blackburn99} (black line).}
\end{figure}

We start moving the cylinder once the flow has settled down to the established 2D shedding regime. A detailed description of the flow is provided in figure \ref{Re500_vorticity}.The five leftmost figures show vorticity contours and  streamlines plotted at five instants spreading over half of the vortex shedding cycle, during which the lift force acts in the upwards direction, starting and ending at times of zero lift. The attachment and separation points are labelled A and S respectively in Figure \ref{Re500_vorticity}(a-e). Comparison with the results of \citet{Blackburn99} shown in the rightmost figures, provides good evidence that the spatial dynamics of the separation bubbles, but also the temporal evolution of the reattachment and separation points is very well predicted. Figure \ref{Re500_displcmt_oscl} shows the evolution of the lift coefficient as a function of the body displacement over the 10 last periods of oscillations. Again, the results are seen to match well the reference data of  \citet{Blackburn99}.


\section{Code scalability}
\label{sec::IBM_scalability}

The aim of the section is to analyze how the (IBM) and in particular the communications between the Eulerian and the Lagrangian spaces affect the scalability of the whole code.  The partitioning strategy for the IBM-\textit{OpenFOAM} method is described hereafter. 

First of all, the Lagrangian markers are stored with the information of the Eulerian mesh and each of them is associated to a \textit{node owner}. At that point, the cell substructure used for the interpolation in the IBM spreading step is created. The starting point is the mesh element containing the Lagrangian marker. The neighbour elements are then systematically checked passing through the faces of the mesh elements of interest. This is done in order to verify that all the elements of the cell structure are in the same partition. When a boundary face is found, three situations may occur:

\begin{enumerate}
\item If the face is defined as a processor boundary (i.e. the boundary of the mesh partition), the \textit{node owner} asks for information to the so-called \textit{ghost point owner}, which is the \textit{node owner} of the neighbour partition.
\item If the face is defined as a periodic boundary, the algorithm looks for the cell through the boundary and a communication \textit{node owner} - \textit{ghost point owner} is established.
\item If another boundary condition is found, the check in that direction is stopped.
\end{enumerate}

\begin{figure}[H]
\begin{center}
\includegraphics[width=0.8\linewidth]{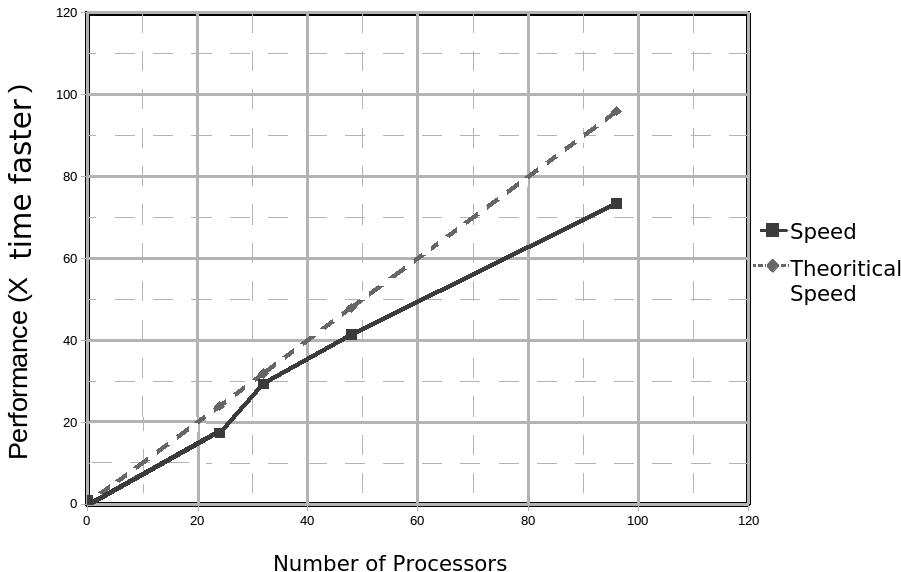}
\end{center}
\caption{ Performance (scability) of the solver as function of the number of processors for a 2D simulation of a flow at Re = 500 past a moving cylinder  using 10$^6$ Eulerian points and 312 Lagrangian markers. \label{fig:scability}}
\end{figure}

Once the value of the parameter $\boldsymbol{\epsilon}$ is calculated for each Lagrangian marker, the force is interpolated using equation (\ref{eq:force1}). Each \textit{node owner} gathers the information on the associated Lagrangian markers, in order to calculate the body force value and it then shares this information with the \textit{ghost owners}. The data are stored as well on a global variable, which is necessary for the resolution of equation (\ref{eq:force1}). The performance of the scalability has been tested and a good efficiency has been observed, as shown in Figure \ref{fig:scability} .

\section{Concluding remarks \& future developments}
\label{sec::conclusions}

The immersed boundary method proposed by \citet{Pinelli20109073} has been implemented within the PISO algorithm of the open source CFD solver \textit{OpenFOAM} for incompressible bluff body fluid flows. The method encompasses the presence of fixed and moving solid obstacles in a computational mesh, without conforming to their boundaries.  Standard  Cartesian  meshes  are  employed,  which  allows  to use efficient and accurate flow solvers. The immersed obstacles are defined using a body force added on the conservation equations, and evaluated on Lagrangian markers that can move over the Eulerian mesh to capture the motion or the deformation of the body. The integration of the method in the finite-volume formalism and in the PISO algorithm has been detailed and a careful verification has been provided using a manufactured solution. The efficiency and the accuracy of the algorithm has been studied on various 2D and 3D simulations of flows around fixed and moving cylinder, including careful comparisons with available numerical and experimental results of the literature. Analysis of the computational cost, numerical behavior and accuracy of the numerical method show that the global properties of the \textit{OpenFOAM} solver are not alterated. A quasi-linear scalability with the number of processors (up to 96) is obtained, with a slope slightly lower than the ideal scalability a feature that has been reported already in existing \textit{OpenFOAM} studies (\citet{OpenFOAM}). 
Work is already in progress to extend the algorithm to the simulation of turbulent flows around bluff bodies, the  main  drawback  of the immersed boundary method being here the difficulty of achieving the desired clustering of grid points toward the obstable walls.

\vspace{1cm}
\section*{Acknowledgements} 
This work was granted access to the HPC resources of Aix-Marseille University financed by the project Equip@Meso (ANR-10-EQPX-29-01). This work was supported by the FUI $N^{\circ}15$ \textit{URABAILA} granted in the frame of the Energy Climate Program of the french government. The financial support of the European Commission through the PELskin FP7 European project (AAT.2012.6.3-1-Breakthrough and emerging technologies) is greatly acknowledged. EC thanks Aix-Marseille University and Direction G\'en\'erale de l'Armement (DGA) for his PhD grant.

\section*{References} 

\bibliographystyle{elsarticle-num-names}

\bibliography{BiblioIBM.bib}

\newpage
\appendix
 \section{PISO Loop implemented in OpenFoam}
\label{subsec::PISO_IBM_implementation_details}

The starting point of the PISO algorithm in OpenFoam is the momentum equation. The left hand side is rewritten as :

\begin{equation}
	\label{eq::NavierStokes_guess_poisson_discretized}
	\frac{\partial u^{\star,1}}{\partial t} + \nabla \cdot (u^{\star,1}u^{\star,1}) - \dfrac{1}{Re} \nabla^2 u^{\star,1}= [UEqnF]	
	\end{equation}
The representation of the equation \ref{eq::NavierStokes_guess_poisson_discretized} within the solver is written as: \textbf{UEqnF} \\
\begin{figure}[H]
\begin{center}
\includegraphics[width=0.45\linewidth]{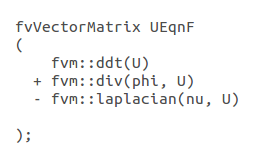}
\end{center}
\end{figure}
The matrix [UEqnF] is then decomposed in 2 matrices :
\begin{equation}
	\label{eq::NavierStokes_A_H_decomposition}
	[UEqnF] = [A] - [H]	
	\end{equation}
with :

\begin{equation}
    [H] = \left(
    \begin{array}{*5{c}}
   
    0 & a_{i;j-1} u^{\star,1}_{i;j-1} & a_{i+1;j-1} u^{\star,1}_{i+1;j-1} \\
    a_{i-1;j} u^{\star,1}_{i-1;j} &  0 &  a_{i+1;j} u^{\star,1}_{i+1;j}\\
    a_{i-1;j+1} u^{\star,1}_{i-1;j+1} &  a_{i;j+1} u^{\star,1}_{i;j+1} &  0 \\

  \end{array}\right )
\end{equation}

\begin{equation}
  [A] = \left( 
    \    \begin{array}{*5{c}}
   
   a_{i-1;j-1} u^{\star,1}_{i-1;j-1} & 0   & 0 \\
    0 &  a_{i;j} u^{\star,1}_{i;j} &  0\\
    0 &  0 &  a_{i+1;j+1} u^{\star,1}_{i+1;j+1} \\

  \end{array}\right ) =  \lbrace a_{ii }\rbrace \times [U^{\star,1}_{ii }]
\end{equation}
In OpenFoam :\\ \\
$[H]$ is the function OpenFoam : \textbf{UEqnF.H()} \\
$\lbrace a_{ii }\rbrace$ is the function OpenFoam : $\textbf{UEqnF.A()}$	\\ \\
The procedure derives the velocity field as :
\begin{equation}
	\label{eq::mat_A_H}
[A] - [H]	= -\nabla p + f(\widehat{u})
	\end{equation}

\begin{equation}
	\label{eq::mat_A}
[A] = \lbrace a_{ii }\rbrace \times [U^{\star,1}_{ii }] 	= -\nabla p + f(\widehat{u}) +  [H]
	\end{equation}

\begin{equation}
	\label{eq::U_corrector_poisson}
[U^{\star,1}_{ii }] 	= \lbrace a^{-1}_{ii }\rbrace (-\nabla p + f(\widehat{u}) +  [H])
	\end{equation}	
where :\\
$\lbrace a^{-1}_{ii }\rbrace$ is written in OpenFoam : $\textbf{rAU}$	\\
$\lbrace a^{-1}_{ii }\rbrace \cdot[H]$ is the vector quantity : \textbf{HbyA} (or \textbf{phiHbyA} if interpolated on the surfaces) \\
 \begin{figure}[H]
\begin{center}
\includegraphics[width=0.6\linewidth]{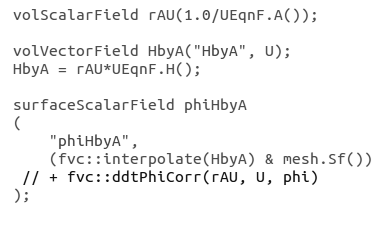}
\end{center}
\end{figure}

Then the Poisson equation is obtained from equation (\ref{eq::U_corrector_poisson}), imposing the continuity equation  (\ref{eq::ConsMass})
	\begin{equation}
	\label{eq::div_expression_poisson}
\nabla \cdot [U^{\star,1}_{ii }] 	= 0 = \nabla \cdot (\lbrace a^{-1}_{ii }\rbrace (-\nabla p + f(\widehat{u})))
	\end{equation}	
	\begin{equation}
	\label{eq::solve_poisson}
\nabla \cdot (\lbrace a^{-1}_{ii }\rbrace \nabla p) = \nabla \cdot (\lbrace a^{-1}_{ii }\rbrace  f(\widehat{u}) ) +  \nabla \cdot (\lbrace a^{-1}_{ii }\rbrace [H])
	\end{equation}
	\ \\
	The equation (\ref{eq::solve_poisson}) is solved in OpenFoam by the command \textit{pEqn.solve()}.
		 \begin{figure}[H]
\begin{center}
\includegraphics[width=0.9\linewidth]{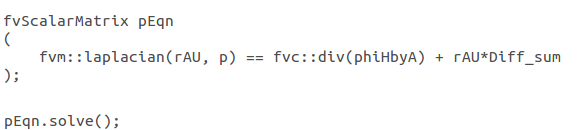}
\end{center}
\end{figure}
The term $ \nabla \cdot \lbrace a^{-1}_{ii }\rbrace  f(\widehat{u})  $ corresponds to the divergence of the force term calculated analytically in OpenFoam  \textbf{Diff\_sum}, as discussed in section \ref{subsec::PISO_modification}. The velocity is then updated, using equation (\ref{eq::U_corrector_poisson}).
\begin{equation*}
[U^{\star,1}_{ii }] 	= \lbrace a^{-1}_{ii }\rbrace (-\nabla p + f(\widehat{u}) +  [H])
	\end{equation*}

In order to update the flux, the force should be interpolated on the surface, introducing the issue discussed in section \ref{subsec::PISO_modification}. Thus, the equation is written as:

	\begin{equation}
	\label{eq::update_flux}
F = S  \cdot [\lbrace a^{-1}_{ii }\rbrace (-\nabla p + f(\widehat{u}) +  [H])]_{faces}
	\end{equation}
\begin{figure}[H]
\begin{center}
\includegraphics[width=0.6\linewidth]{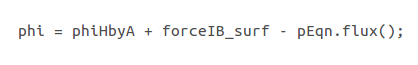}
\end{center}
\end{figure}
with $f(\widehat{u})_{faces}$ the immersed boundary force calculated analytically on the surface.

\subsection{Algorithm}
\label{subsec::Solver_scheme}

The time step $n+1$ can be thus represented by the following diagram:
\newpage
\begin{figure}[H]
\begin{center}
\includegraphics[width=1\linewidth]{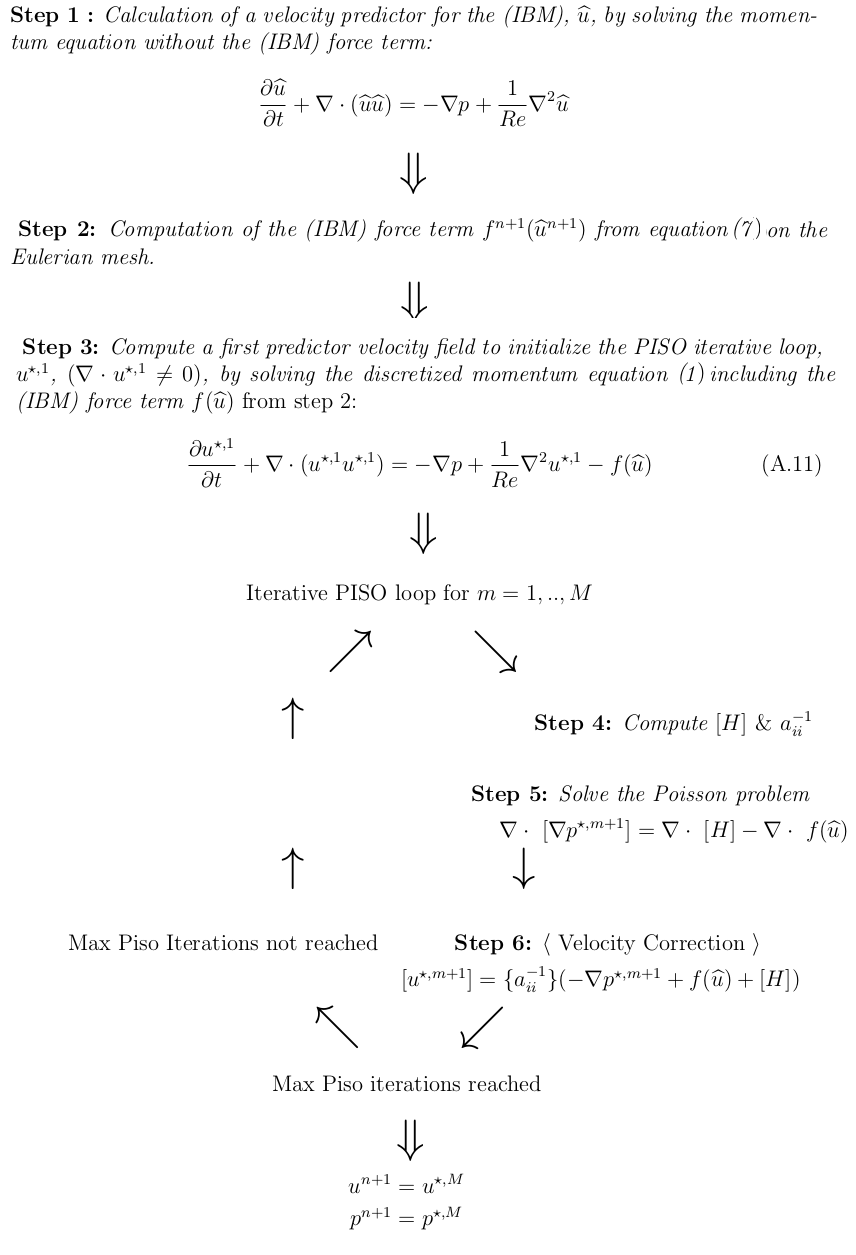}
\end{center}
\end{figure}
\pagenumbering{gobble}

\end{document}